\documentclass[11pt]{article}
\usepackage{amsmath}
\usepackage{amsfonts}
\usepackage{amssymb}
\usepackage{graphicx}
\usepackage{url}%

\begin{document}

\title{Indeterminacy of Spatiotemporal Cardiac Alternans}
\author{Xiaopeng Zhao \\ Mechanical, Aerospace and Biomedical Engineering Department \\
University of Tennessee \\
Knoxville, TN 37996\\
xzhao9@utk.edu}
\date{\today}
\maketitle

\begin{abstract}
Cardiac alternans, a beat-to-beat alternation in action potential
duration (at the cellular level) or in ECG morphology (at the whole
heart level), is a marker of ventricular fibrillation, a fatal heart
rhythm that kills hundreds of thousands of people in the US each
year. Investigating cardiac alternans may lead to a better
understanding of the mechanisms of cardiac arrhythmias and
eventually better algorithms for the prediction and prevention of
such dreadful diseases. In paced cardiac tissue, alternans develops
under increasingly shorter pacing period. Existing experimental and
theoretical studies adopt the assumption that alternans in
homogeneous cardiac tissue is exclusively determined by the pacing
period. In contrast, we find that, when calcium-driven alternans
develops in cardiac fibers, it may take different spatiotemporal
patterns depending on the pacing history. Because there coexist
multiple alternans solutions for a given pacing period, the
alternans pattern on a fiber becomes unpredictable. Using numerical
simulation and theoretical analysis, we show that the coexistence of
multiple alternans patterns is induced by the interaction between
electrotonic coupling and an instability in calcium cycling.
\end{abstract}

\section{Introduction}

\begin{figure}[tbh]
\centering\includegraphics[width=2.5in]{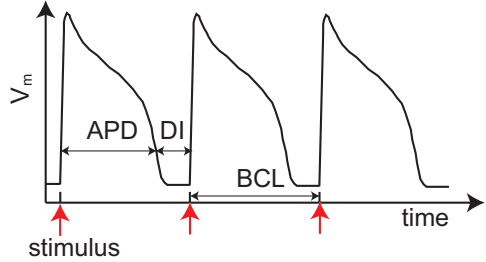}\caption{Schematic
action potential showing the response of the transmembrane voltage to periodic
electrical stimuli.}%
\label{fig:action_potential}%
\end{figure}

Sudden cardiac death, attributable to unexpected ventricular arrhythmias, is
one of the leading causes of death in the US and kills over 300,000 Americans
each year \cite{AHA}. The induction and maintenance of ventricular arrhythmias
has been linked to single-cell dynamics \cite{Karma1994,Garfinkel2000}. In
response to an electrical stimulus, cardiac cells fire an action potential
\cite{Plonsey2000}, which consists of a rapid depolarization of the
transmembrane voltage (V$_{\text{m}}$) followed by a much slower
repolarization process before returning to the resting value (Fig.
\ref{fig:action_potential}). The time interval during which the voltage is
elevated is called the action potential duration (APD). The time between the
end of an APD and the beginning of the next one is called the diastolic
interval (DI). The time interval between two consecutive stimuli is called the
basic cycle length (BCL). When the pacing rate is slow, a periodic train of
electrical stimuli produces a phase-locked steady-state response, where each
stimulus gives rise to an identical action potential (1:1 pattern). When the
pacing rate becomes sufficiently fast, the 1:1 pattern may be replaced by a
2:2 pattern, so-called electrical alternans \cite{Nolasco1968JAP,Panfilov1998}%
, where the APD alternates between short and long values. Recent experiments
have established a causal link between alternans and the risk for ventricular
arrhythmias \cite{Pastore1999Circulation,Bloomfield2006ACC,shusterman2006Cir}.
Therefore, understanding mechanism of alternans is a crucial step in detection
and prevention of fatal arrhythmias.

\begin{figure}[tbh]
\centering
\includegraphics[width=2.5in]{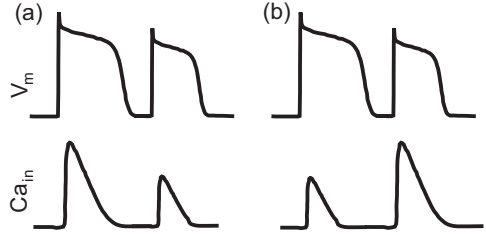}\caption{Schematic
illustration of electromechanically in-phase (a) and out-of-phase (b)
alternans.}%
\label{fig:EM_coupling}%
\end{figure}

Cellular mechanisms of alternans have been much studied. Summaries on this
topic can be found in recent review articles by Shiferaw et al.
\cite{Shiferaw2006Review} and Weiss et al. \cite{Weiss2006CircRes}. At the
cellular level, cardiac dynamics involves bidirectional coupling between
membrane voltage (V$_{\text{m}}$) dynamics and intracellular calcium
(Ca$_{\text{in}}$) cycling. During an action potential, the elevation of
V$_{\text{m}}$ activates L-type Ca currents $I_{\text{Ca}}$ to invoke the
elevation of [Ca$_{\text{in}}$], which in turn triggers Ca release from the
sarcoplasmic reticulum (SR), a procedure known as calcium-induced-calcium
release (CICR) \cite{EndoPhysiol1977}. The V$_{\text{m}}\rightarrow
$Ca$_{\text{in}}$ coupling satisfies graded release, where a larger DI leads
to an increase in the Ca release at the following beat since it allows more
time for L-type Ca channels to recover. On the other hand, Ca release from the
SR affects the APD in two folds: to curtail the APD by enhancing the
inactivation of L-type Ca currents $I_{Ca}$; and to prolong the APD by
intensifying Na$^{+}$/Ca$^{2+}$ exchange currents $I_{\text{NaCa}}$.
Therefore, depending on the relative contributions of $I_{\text{Ca}}$ and
$I_{\text{NaCa}}$, an increase in Ca release may either shorten the APD
(negative Ca$_{\text{in}}\rightarrow$V$_{\text{m}}$ coupling) or lengthen the
APD (positive Ca$_{\text{in}}\rightarrow$V$_{\text{m}}$ coupling)
\cite{Shiferaw2005PRE,Shiferaw2006PNAS,Sato2006CircRes}. There exist two main
cellular mechanisms of alternans. Firstly, alternans may be attributed to
steep APD restitution \cite{Nolasco1968JAP}, which is due to a period-doubling
instability in the V$_{\text{m}}$ dynamics. In this case, Ca$_{\text{in}}$
transient alternans, as a slave variable, is induced because V$_{\text{m}}$
regulates [Ca$_{\text{in}}$] via the L-type Ca currents and the sodium-calcium
exchange currents. Secondly, alternans may be caused by a period-doubling
instability in Ca$_{\text{in}}$ cycling, which is associated with a steep
relationship between the sarcoplasm reticulum (SR) release and SR load
\cite{Chudin1999Biophys,Diaz2004CircRes}. In this case, APD alternans is a
secondary effect via Ca$_{\text{in}}\rightarrow$V$_{\text{m}}$ coupling. For
ease of reference, we call the first mechanism \emph{APD-driven alternans} and
the second \emph{Ca-driven alternans}. Interestingly, APD and Ca$_{\text{in}}$
transient alternans can be electromechanically (E/M) in phase or out of phase
\cite{Rubenstein1995Circ,Walker2003CardioRes}. In E/M in-phase alternans, a
long-short-long APD pattern corresponds to a large-small-large [Ca$_{\text{in}%
}$] pattern, See Fig. \ref{fig:EM_coupling} (a). In contrast, in E/M
out-of-phase alternans, a long-short-long APD pattern corresponds to a
small-large-small [Ca$_{\text{in}}$] pattern, See Fig. \ref{fig:EM_coupling}
(b). When alternans happens in isolated cells, the bidirectional coupling
between APD and Ca$_{\text{in}}$ transient determines the relative phase of
APD and Ca$_{\text{in}}$ transient alternans. In particular, APD-driven
alternans always leads to E/M in-phase alternans whereas Ca-driven alternans
is E/M in phase for positive Ca$_{\text{in}}\rightarrow$V$_{\text{m}}$
coupling and out of phase for negative Ca$_{\text{in}}\rightarrow$%
V$_{\text{m}}$ coupling
\cite{Shiferaw2005PRE,Shiferaw2006PNAS,Sato2006CircRes}. 

The mechanism of alternans in multicellular tissue is more complicated since
it involves electrotonic coupling and conduction velocity restitution. Of
particular interest is a phenomenon called spatially discordant alternans, in
which different regions of the tissue alternate out of phase. Discordant
alternans is arrhythmogenic because it forms a dynamically heterogeneous
substrate that may promote wave break and reentry
\cite{Karma1994,Garfinkel2000}. To study the spatiotemporal patterns of
alternans, Echebarria and Karma derived amplitude equations that are based on
APD-driven alternans \cite{Echebarria2002PRL,Echebarria2006}. These amplitude
equations not only are capable of quantitative predictions but also provide
insightful understandings on the arrhythmogenic patterns. In a recent article,
Dai and Schaeffer \cite{Dai2008} analytically computed the linear spectrum of
Echebarria and Karma's amplitude equations for the cases of small dispersion
and long fibers.

Spatial patterns of alternans have been investigated in experiments
\cite{Pastore1999Circulation,Choi2000JPhysiol,Pruvot2004CircRes,Katra2004HeartCircPhy}%
. Recently, Aistrup et al. \cite{Aistrup2006CircRes} used single-photon
laser-scanning confocal microscopy to measure Ca signaling in individual
myocytes. They found that Ca alternans is spatially synchronized at low pacing
rates whereas dyssynchronous patterns, where a number of cells are out of
phase with adjoining cells, arise when the pacing rate increases. Aistrup et
al. also observed subcellular alternans at fast pacing, where Ca alternans is
spatially dyssynchronous within a cell. Using simulations of 1-d homogeneous
tissue, Sato et al. \cite{Sato2006CircRes} found that, in cardiac fibers with
negative Ca$\rightarrow$V$_{\text{m}}$ coupling, Ca alternans reverses phase
over a length scale of one cell whereas, in fibers with positive
Ca$\rightarrow$V$_{\text{m}}$ coupling, Ca alternans changes phase over a much
larger length scale. They interpreted this difference by showing that negative
Ca$\rightarrow$V$_{\text{m}}$ coupling tends to desynchronize two coupled
cells while positive Ca$\rightarrow$V$_{\text{m}}$ coupling tends to
synchronize the coupled cells.

Motivated by the aforementioned experimental and theoretical work, this paper
aims to explore spatial patterns of cardiac alternans. Through extensive
numerical simulations, we find that complex spatial patterns of Ca alternans
with phase reversals in adjacent cells can happen in homogeneous fibers with
both negative and positive Ca$\rightarrow$V$_{\text{m}}$ couplings. Most
surprisingly, we find that the spatiotemporal pattern of cardiac alternans is
not determined by the pacing period alone. Specifically, when calcium-driven
alternans develops in multicellular tissue, there coexist multiple
spatiotemporal patterns of alternans regardless of the length of the fiber,
the junctional diffusion of Ca, and the type of Ca$\rightarrow$V$_{\text{m}}$
coupling. We further investigate the mechanism that leads to the coexistence
of multiple alternans solutions. Our analysis shows that multiple alternans
solutions are induced because of the interaction between electrotonic coupling
and an instability in Ca$_{\text{in}}$ cycling.

\section{Model Description}

\subsection{Membrane dynamics}

\begin{figure}[tbh]
\centering
\begin{tabular}
[c]{cc}%
\includegraphics[width=2.25in]{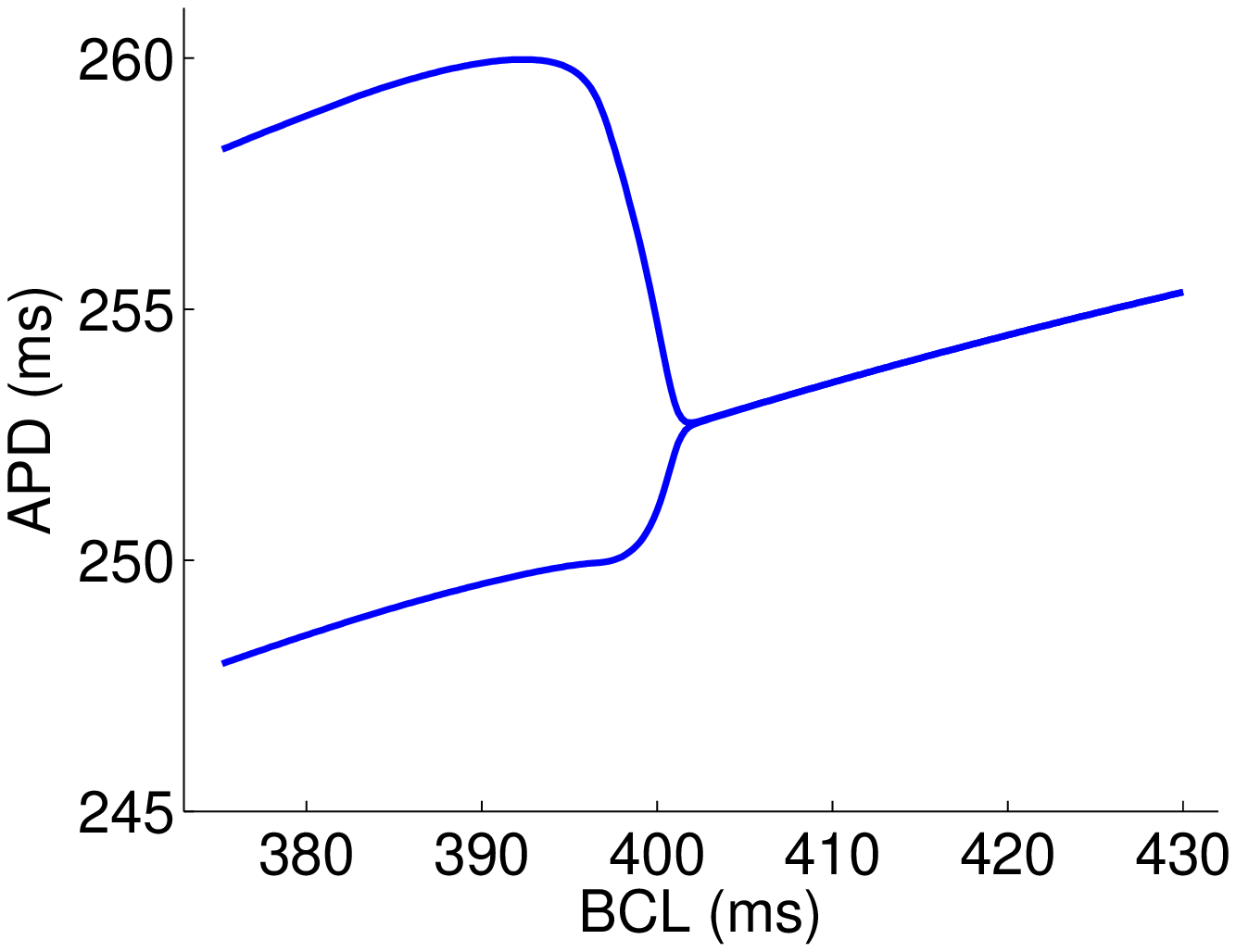} &
\includegraphics[width=2.25in]{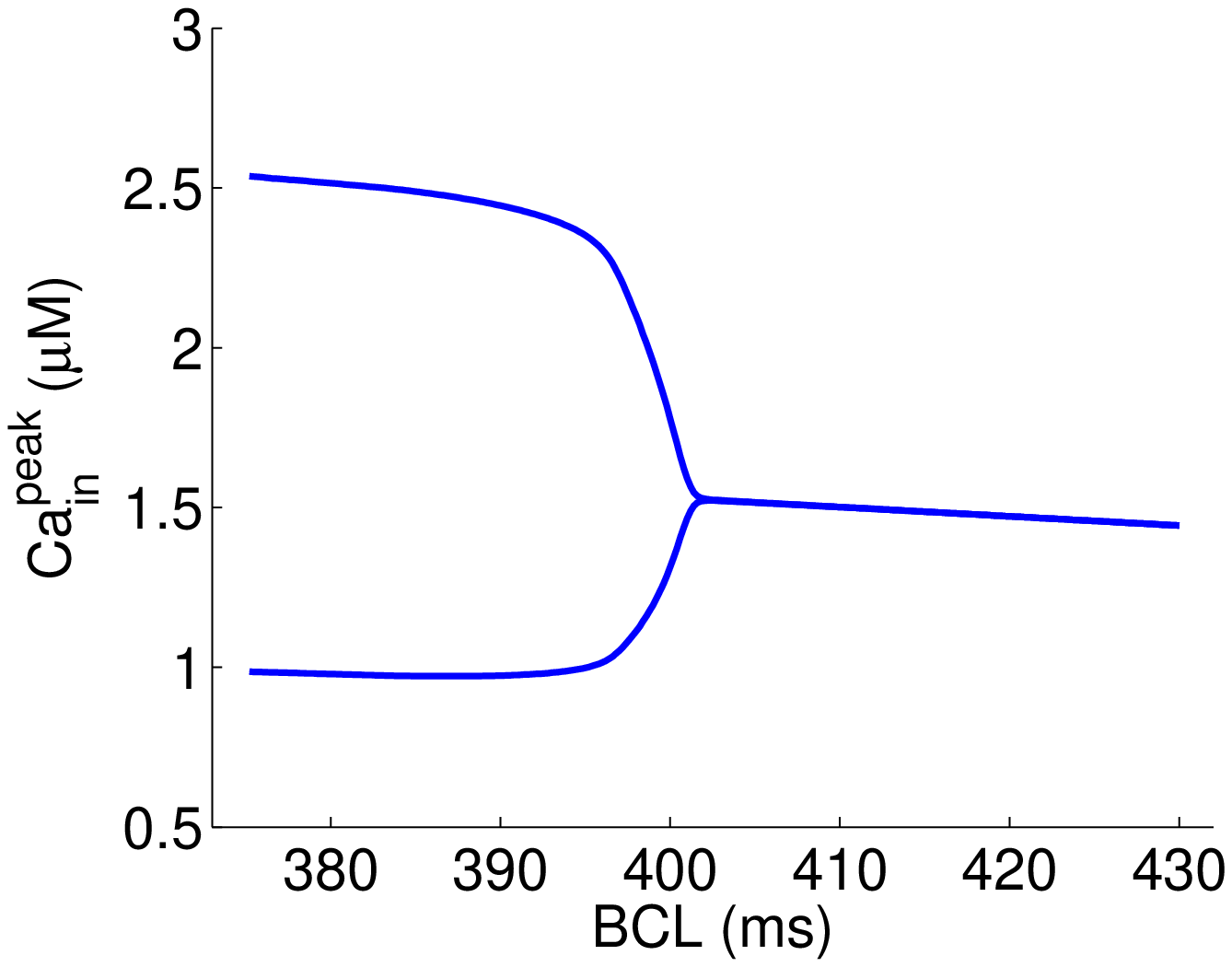}\\
(a) & (b)
\end{tabular}
\caption{Bifurcation diagrams of an isolated cell using the Shiferaw-Fox
model.}%
\label{fig:ode_bif}%
\end{figure}

We adopt a model of membrane dynamics that combines the calcium dynamics model
developed by Shiferaw et al. \cite{Shiferaw2003BioPhy} and the canine ionic
model by Fox et al. \cite{Fox2002AmJP}. In the following, we will refer to
this model as the Shiferaw-Fox model. Detailed formulations of the model can
be found in \cite{Shiferaw2003BioPhy,ShiferawCode}. The Shiferaw-Fox model has
adopted two sets of parameters in the calcium dynamics to account for negative
and positive Ca$_{\text{in}}\rightarrow$V$_{\text{m}}$ couplings. Besides the
phase difference, the two sets of parameters also produce alternans at
different values of BCL. Using Shiferaw's default parameters, we find
alternans happens at BCL$\approx401$ ms for negative Ca$_{\text{in}%
}\rightarrow$V$_{\text{m}}$ coupling and BCL$\approx323$ ms for positive
Ca$_{\text{in}}\rightarrow$V$_{\text{m}}$ coupling. Fig. \ref{fig:ode_bif}
shows the bifurcation diagrams in APD and peak value of [Ca$_{\text{in}}$] for
negative Ca$_{\text{in}}\rightarrow$V$_{\text{m}}$ coupling. The bifurcation
diagrams for positive Ca$_{\text{in}}\rightarrow$V$_{\text{m}}$ coupling are
similar and thus are not shown here. We note that, in simulations of isolated
cells using the Shiferaw-Fox model, alternans solutions do not depend on the
initial condition nor on the pacing history. However, as we will show in the
following, fibers based on the Shiferaw-Fox model possess multiple alternans
solutions, which are sensitive to the initial condition and the pacing protocol.

\subsection{Simulation of fiber models}

We study paced, homogeneous fibers, which can be modeled using the cable
equation:%
\begin{equation}
\frac{\partial v}{\partial t}=D\,\frac{\partial^{2}v}{\partial x^{2}}-\frac
{1}{C_{m}}\left(  I_{\text{ion}}+I_{\text{ext}}\right)  , \label{eqn:voltage}%
\end{equation}
where $v$ represents V$_{\text{m}}$, $D=5\times10^{-4}$ cm$^{2}/$ms represents
the effective diffusion coefficient of $v$ in the fiber, $C_{m}=1$ $\mu
F/$cm$^{2}$ represents the transmembrane capacitance, $I_{\text{ion}}$ is the
total ionic current, and $I_{\text{ext}}$ represents the external current
stimulus. The ionic current $I_{\text{ion}}$ is computed using the
Shiferaw-Fox model. The current stimulus has duration 1 ms and amplitude 80
$\mu$A/$\mu$F. This paper studies fibers of various lengths. For two coupled
cells, we pace the left cell. For longer fibers, we pace the leftmost few
cells to ensure propagation. For example, the leftmost 5 cells are paced in
simulating a fiber of 100 cells. The cable equation (\ref{eqn:voltage}) is
solved using the finite difference method with a space step of $\Delta
x=0.015$ cm and time step of $\Delta t=0.1$ ms. No-flux boundary conditions
are imposed at both ends of the fiber \cite{Shiferaw2003BioPhy,ShiferawCode}.

\subsection{Pacing protocols}

To study the onset and development of alternans, we pace both single cells and
fibers of various lengths with several pacing protocols, which are briefly
described below.

(i) In the \emph{downsweep protocol} \cite{Kalb2004JCardiovasc}, the
cell/fiber is paced periodically with period BCL until it reaches steady
state. Then, the pacing period is reduced by $\Delta$B and the procedure is
repeated many times. Note that this protocol is also known as \emph{dynamic
pacing protocol} \cite{Riccio1999CircRes}.

(ii) The \emph{perturbed downsweep protocol}, proposed by Kalb et al.
\cite{Kalb2004JCardiovasc}, can be regarded as a perturbation to the downsweep
protocol. At each pacing period BCL, the cell/fiber is first paced N beats to
reach steady state. Then, a longer pacing period is applied at the N+1st
pacing, after which the original pacing period is applied for 10 beats to
allow the tissue to recover its previous steady state. Next, a shorter pacing
period is applied and followed by 10 beats of the original pacing period.
Finally, the pacing period is reduced by $\Delta$B and the procedure is repeated.

(iii) To explore the possibility for multiple alternans solutions, we set up
certain initial condition and pace the tissue with period BCL to reach steady
state, a process we call \emph{direct pacing}.

(iv) To explore the origin of an alternans pattern, we use the \emph{upsweep
protocol} \cite{Kalb2004JCardiovasc}, which is a reversed downsweep protocol.

\section{Spatiotemporal Patterns of Alternans: Numerical Exploration}

We simulate fibers using the Shiferaw-Fox model with both negative and
positive Ca$_{\text{in}}\rightarrow$V$_{\text{m}}$ couplings under various
conditions. Default parameters in Shiferaw's code \cite{ShiferawCode} are used
unless otherwise specified. Despite quantitatively significant differences, we
find both types of couplings lead to the coexistence of multiple alternans
solutions. For clarity, we start with the results for negative Ca$_{\text{in}%
}\rightarrow$V$_{\text{m}}$ coupling and defer the results for positive
Ca$_{\text{in}}\rightarrow$V$_{\text{m}}$ coupling in a later subsection.

\subsection{Coexistence of multiple solutions}

\begin{figure}[tbh]
\centering%
\begin{tabular}{ccc}
\includegraphics[width=2.in]{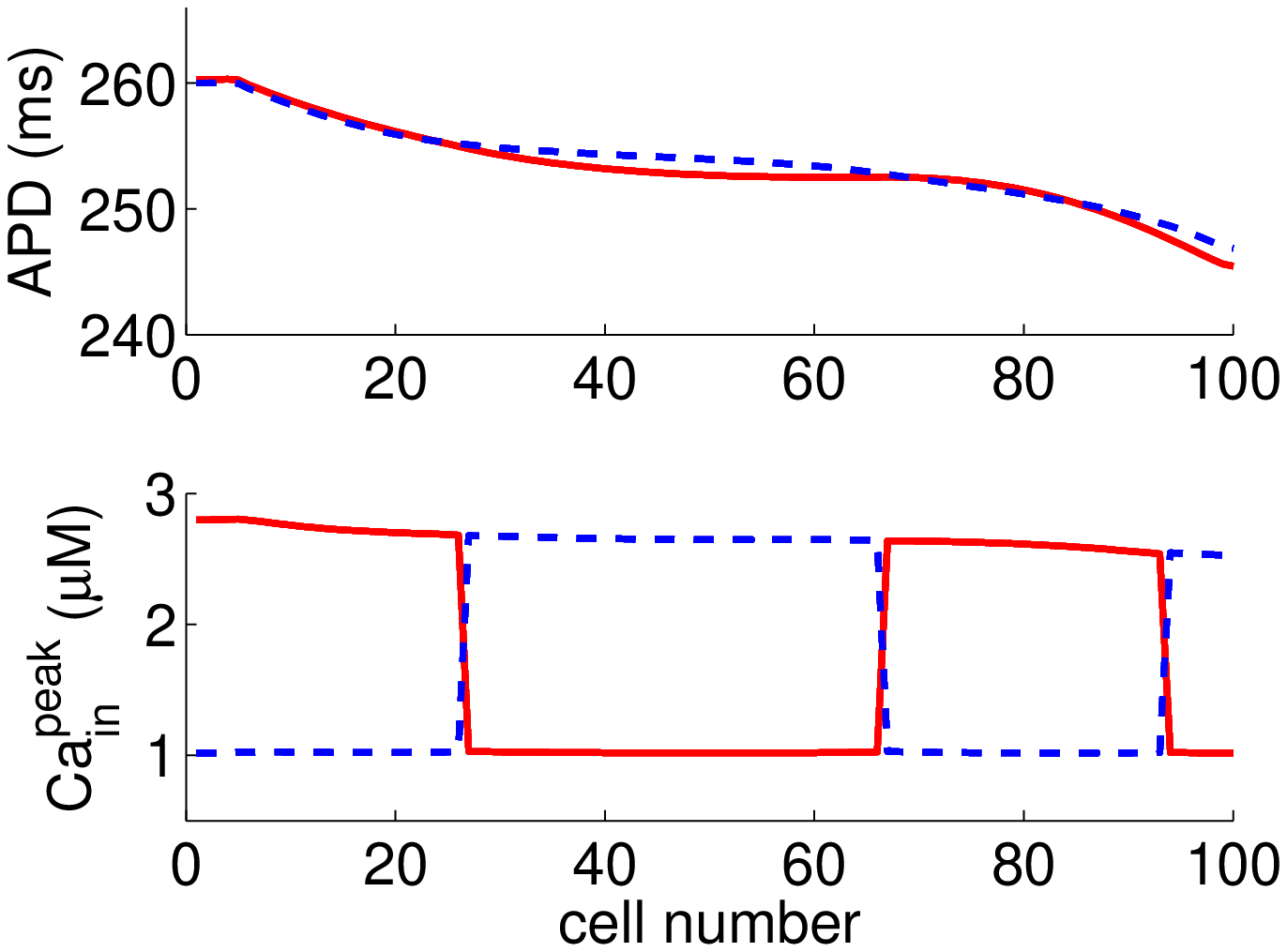} & %
\includegraphics[width=2.in]{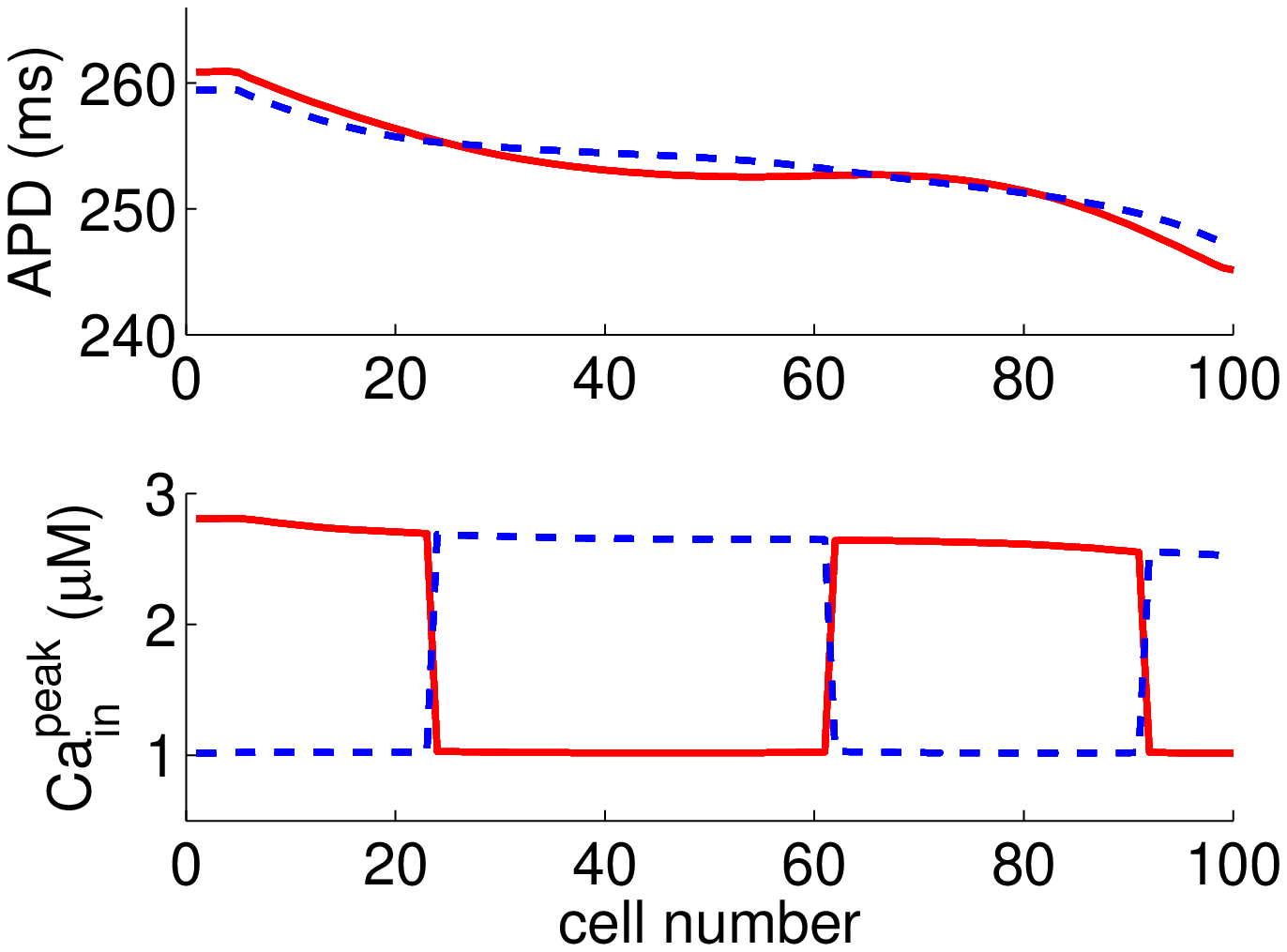} & %
\includegraphics[width=2.in]{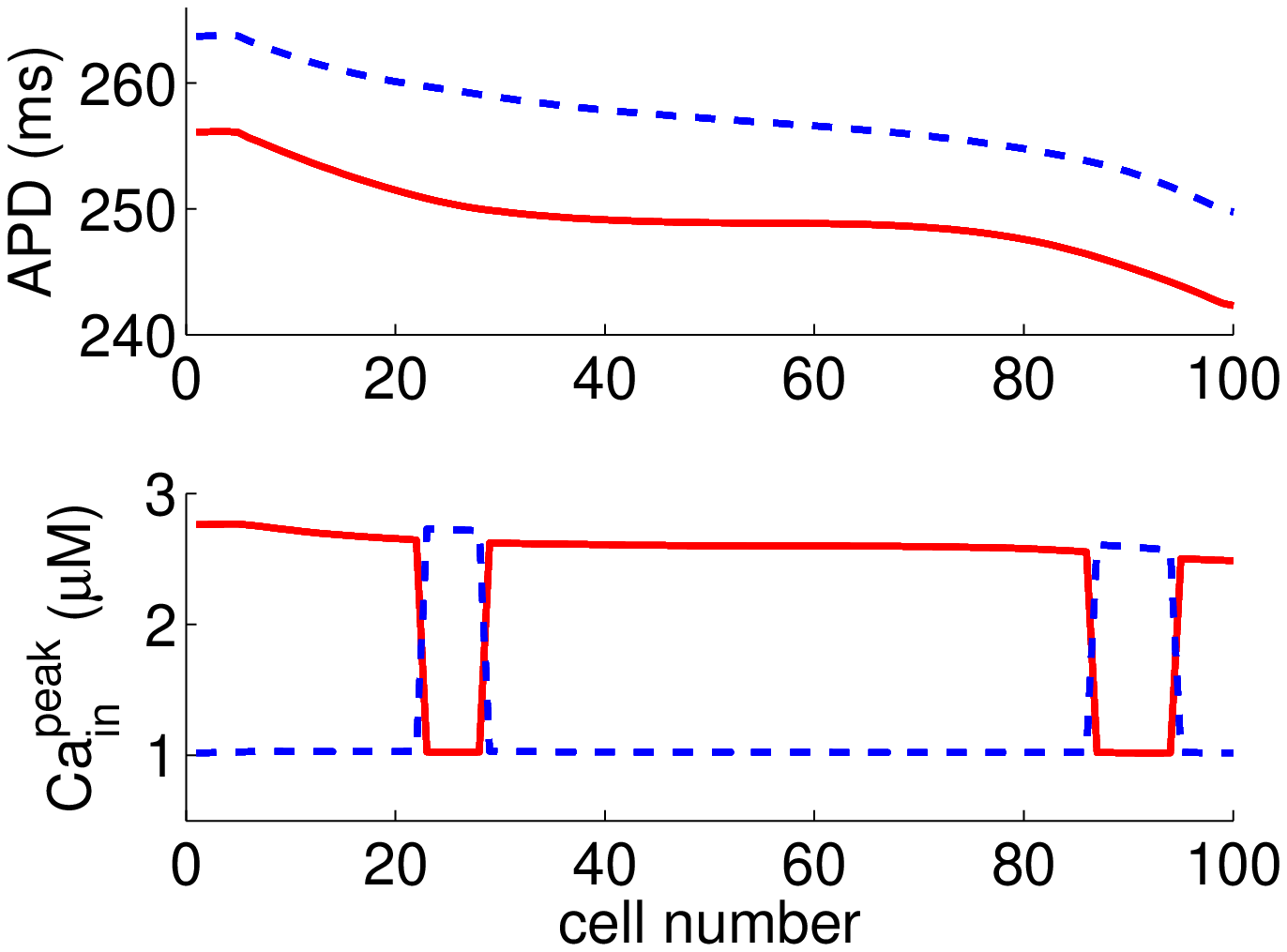} \\
(a) 500:-1:375 (ms) & (b) 499:-2:375 (ms) & (c) 500:-25:375 (ms) \\
\includegraphics[width=2.in]{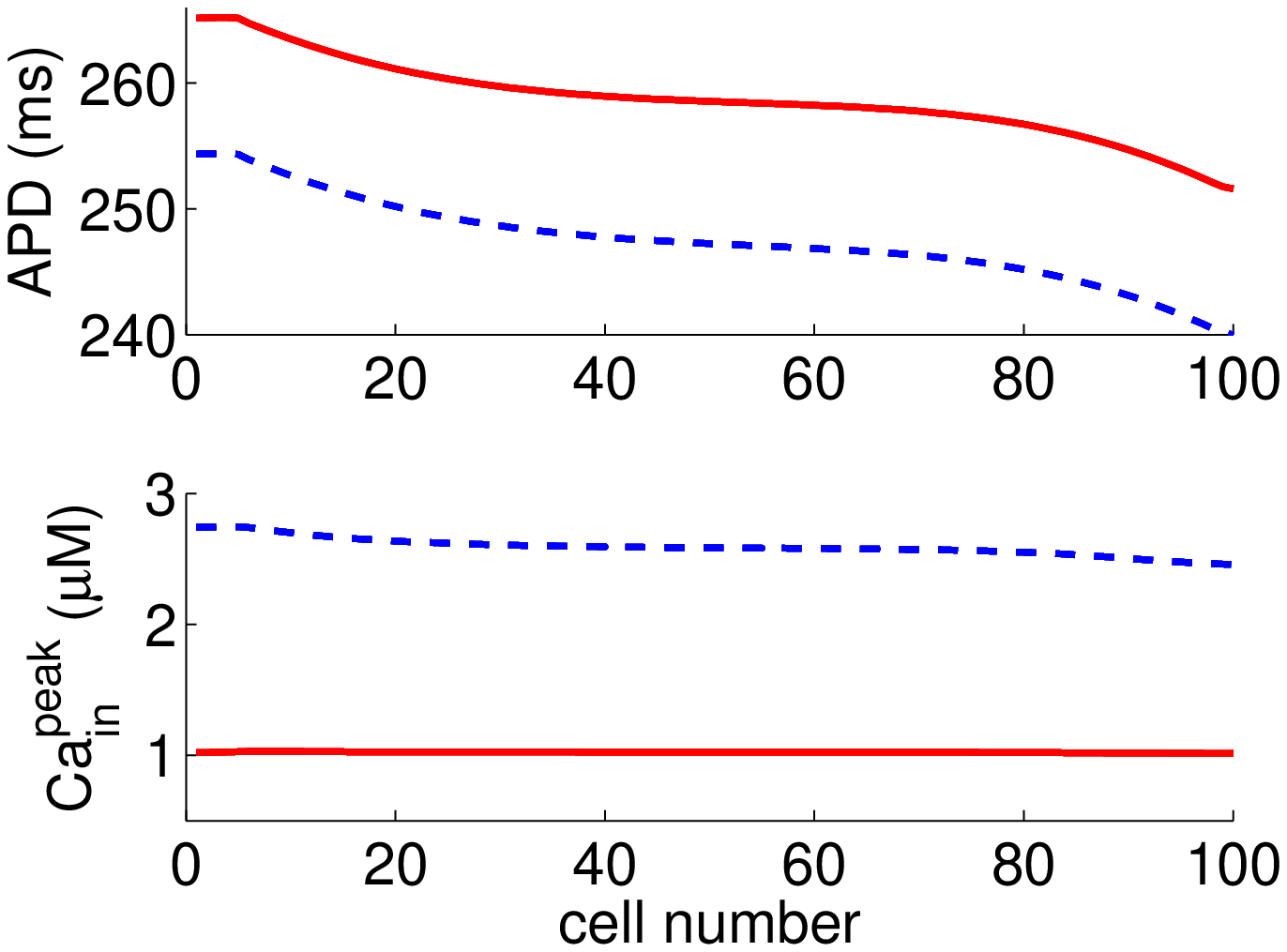} & %
\includegraphics[width=2.in]{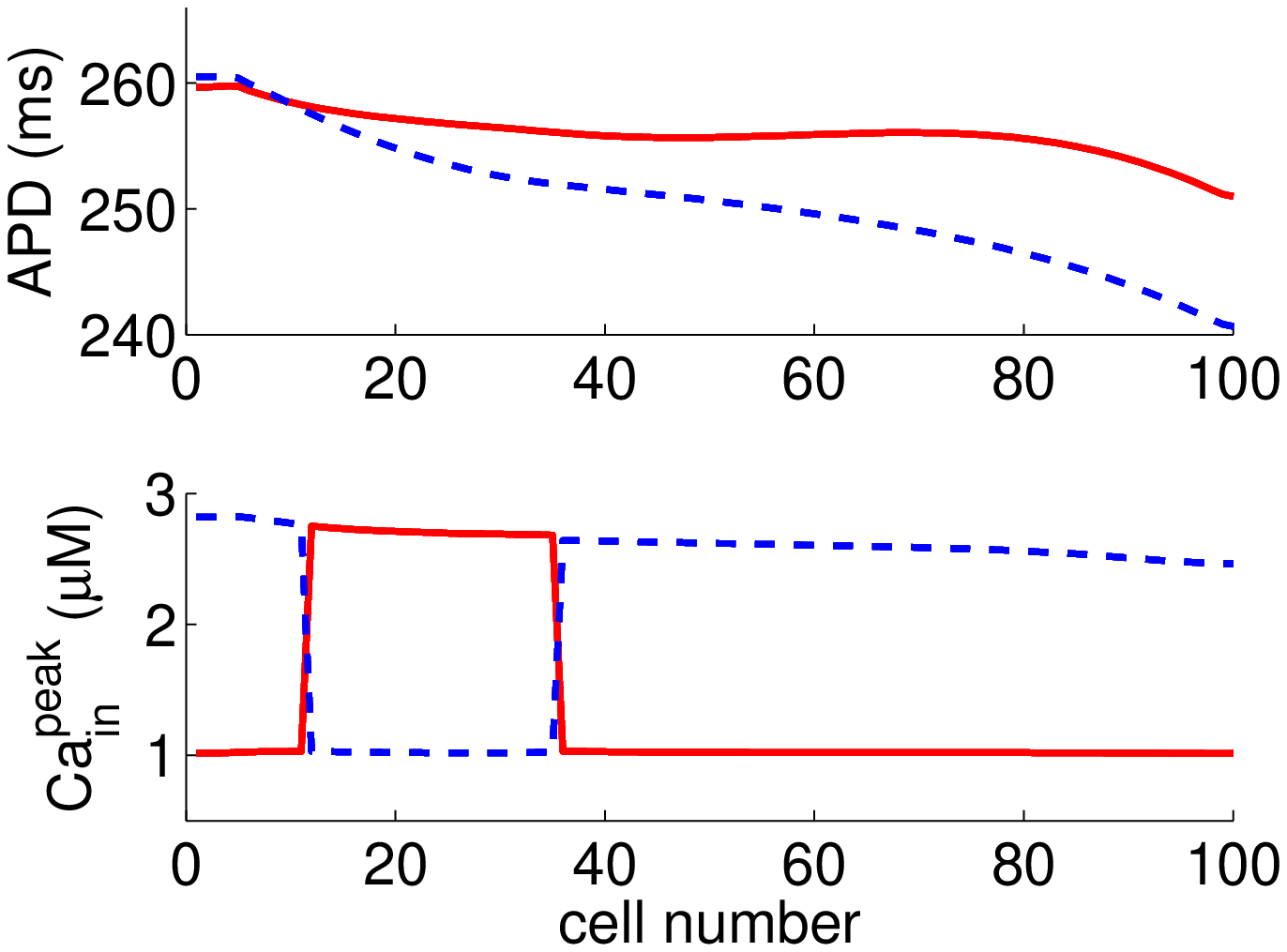} & %
\includegraphics[width=2.in]{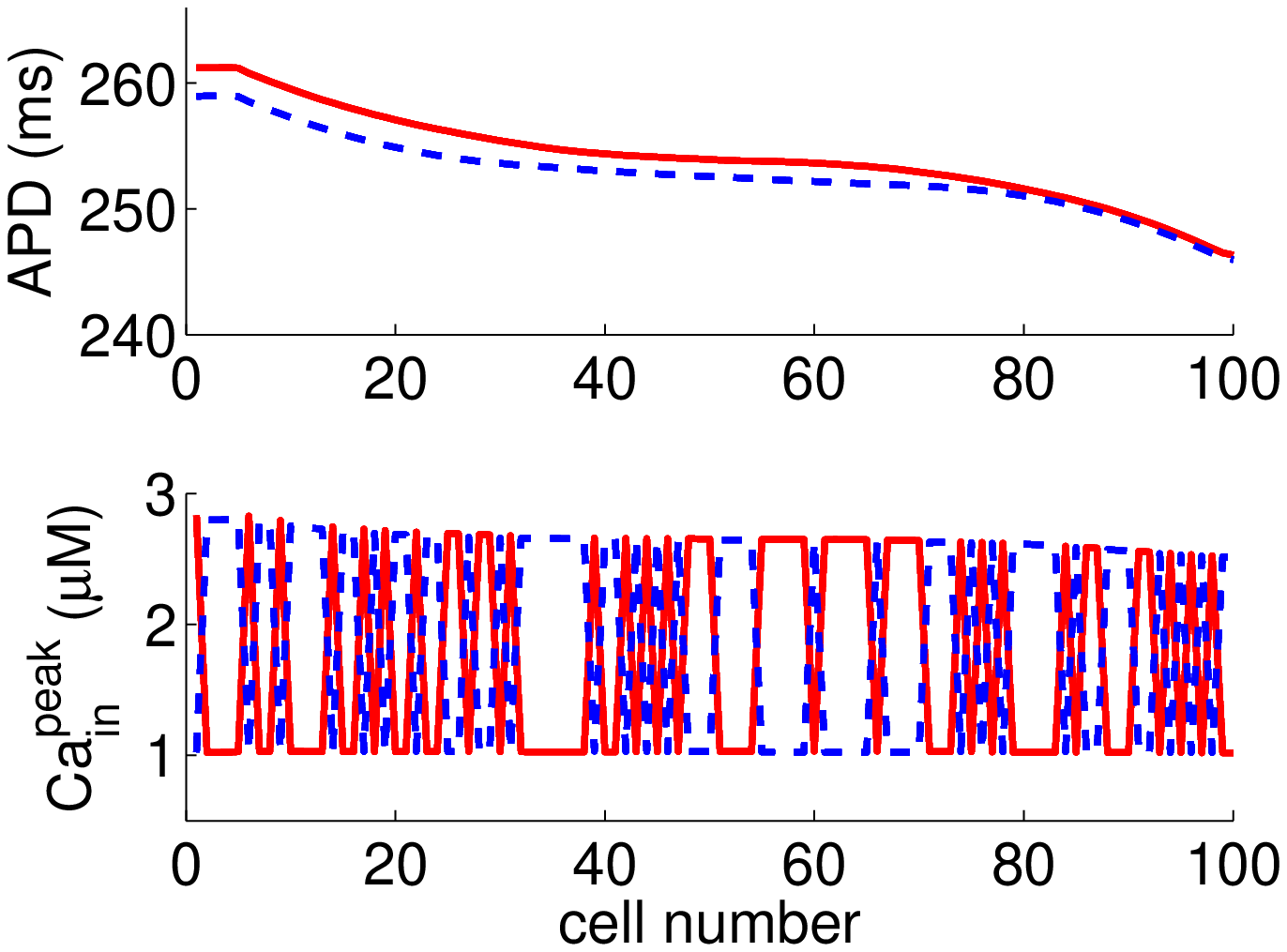} \\
(d) uniform distribution & (e) uneven distribution & (f) random distribution%
\end{tabular}%
\caption{Six selected alternans solutions for a homogeneous fiber of
100 cells with negative Ca$\rightarrow V_{m}$ coupling when paced at
BCL=$375$ ms. Panels (a)-(c) are obtained using downsweep protocols
from $500$ ms to $375$ ms in steps of $1$, $2$, and $25$ ms,
respectively. Panels (d)-(f) are obtained by directly pacing at
BCL=$375$ ms with different initial distributions in
Ca$_{\text{in}}$, see text for details. To verify if a solution
reaches steady state, we plot the last 10 beats of the simulation
results, where odd beats are represented by red solid lines and even
beats by blue dashed lines. The coincidence of all odd beats and
that of all even beats confirm that the solution is indeed in steady
state.} \label{fig:Dtype100}
\end{figure}

We first consider a homogeneous fiber of 100 cells with negative
Ca$_{\text{in}}\rightarrow$V$_{\text{m}}$ coupling. To our surprise, numerical
simulations show that when the fiber is in alternans, there coexist multiple
solutions for a given pacing period. For example, Fig.~\ref{fig:Dtype100}
shows 6 selected solutions of alternans for the fiber paced at BCL=$375$~ms.
Here, the steady-state solutions in panels (a-c) are obtained using the
downsweep protocol with step size $\Delta$B=1 ms, 2 ms, and 25 ms,
respectively. The pacing protocols are started from BCL=500 ms in (a) and (c)
and from BCL=499 ms in (b). We note that the solution of a downsweep protocol
is not influenced by the initial condition at the starting, long BCL; instead,
the solution is sensitive to the step size $\Delta$B. The steady-state
solutions in panels (d-f) are obtained by pacing the fiber at BCL=$375$~ms
with prescribed initial conditions for $200$ beats, the so-called direct
pacing. The initial condition of the fiber in panel (d) is uniform, i.e., all
cells are assigned the same resting voltage, gating variables, and ionic
concentrations. The initial condition in panel (e) is same as that in (d)
except that [Ca$_{\text{in}}$] is assigned to be $0.54$ $\mu$M for the first
35 cells and $0.66$ $\mu$M for the remaining cells. Interestingly, this
initial condition leads to a steady-state [Ca$_{\text{in}}$] pattern, which,
besides the phase reversal between cells 35 and 36, has another phase reversal
between cells 11 and 12. The initial condition in panel (f) is the same as
that in (d) except that [Ca$_{\text{in}}$] is randomly assigned for cells on
the fiber according to a uniform distribution in the interval of $0.45$ $\mu$M
to $0.75$ $\mu$M. In all protocols, we pace the fiber for 200 beats at each
BCL and plot the last 10 beats at BCL=375 ms. The simulation results in Fig.
\ref{fig:Dtype100} demonstrate that the alternans on a fiber is not solely
determined by the pacing period. Instead, the solution is sensitive to the
pacing protocol and the initial condition. It is worth noting that, besides
the solutions shown in Fig. \ref{fig:Dtype100}, there exist many other
solution patterns. In particular, there exist many complex patterns similar to
Fig. \ref{fig:Dtype100} (f). In the following, we will verify whether the
phenomenon is influenced by the length of the fiber, junctional Ca diffusion,
or Ca$_{\text{in}}\rightarrow$V$_{\text{m}}$ coupling.

\begin{figure}[tbh]
\centering
\begin{tabular}{cc}
\includegraphics[width=2.in]{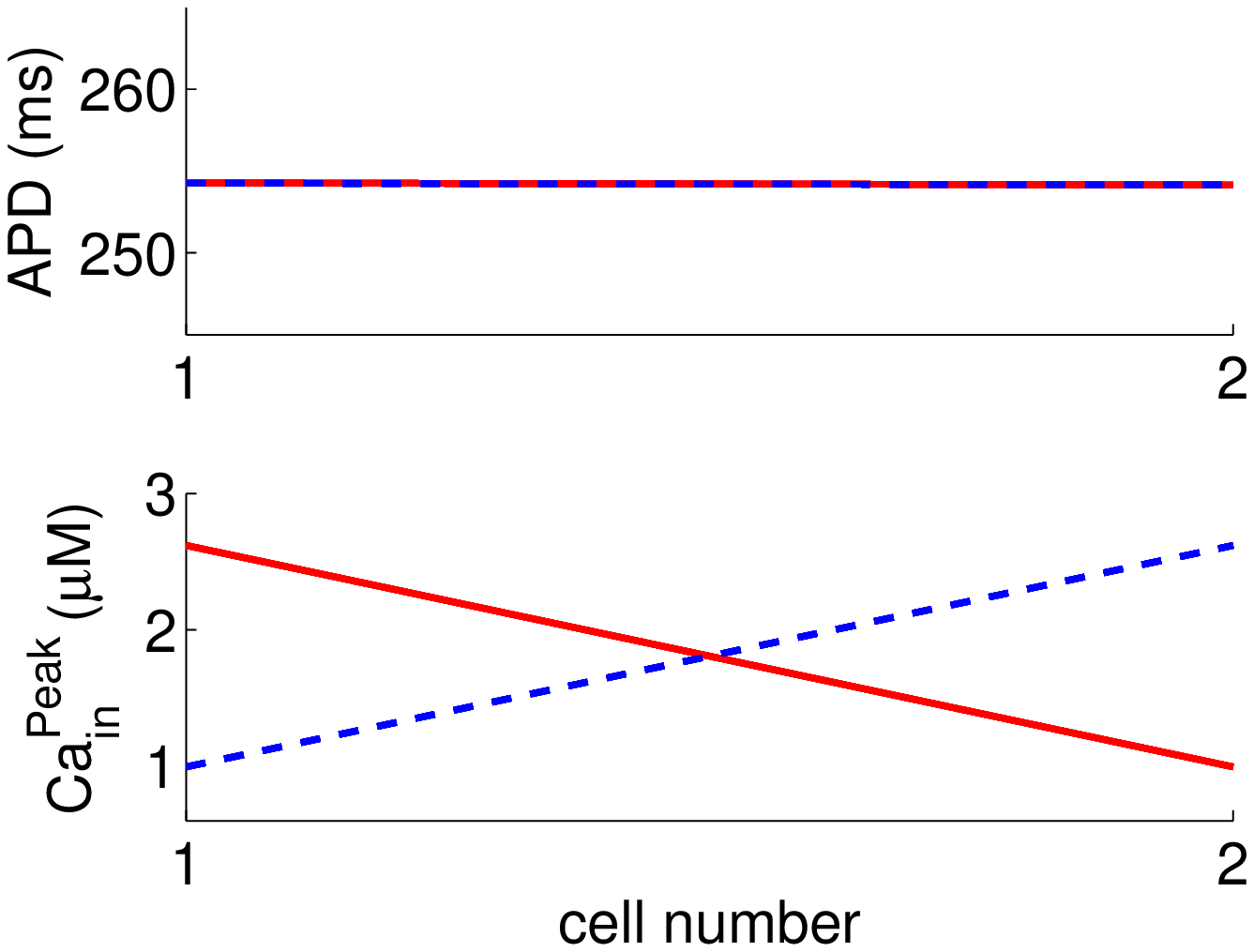} & %
\includegraphics[width=2.in]{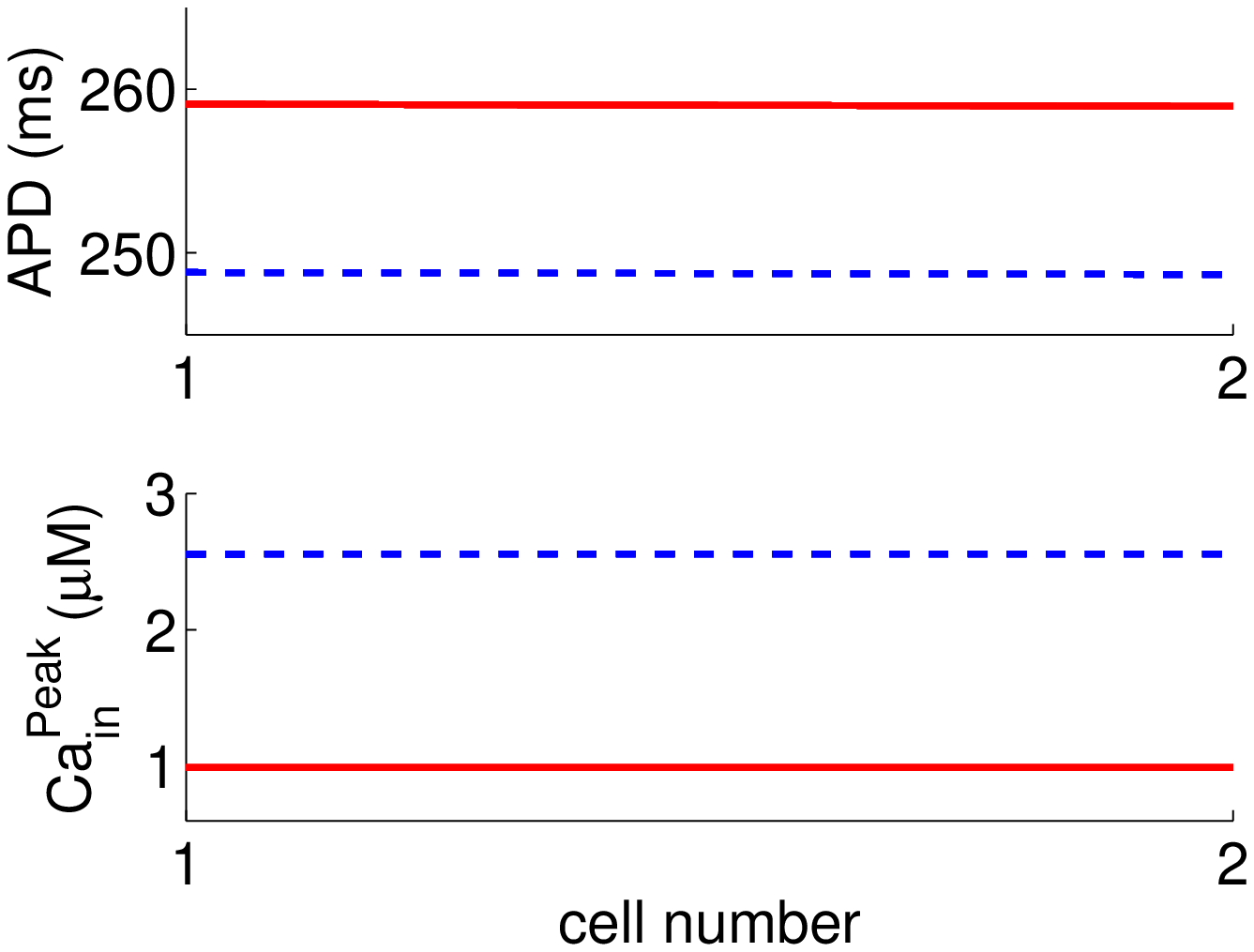} \\
(a) & (b)%
\end{tabular}%
\caption{A ``fiber'' of two cells with negative Ca$\rightarrow
V_{m}$ coupling can have both spatially desynchronized and
synchronized alternans solutions. The pacing period is $375$~ms. The
last 10 beats of the steady-state solutions are plotted, where odd
beats are represented by red solid lines and even beats by blue
dashed lines. } \label{fig:Dtype2}
\end{figure}

\subsection{Influence of the length of the fiber}

We repeat the numerical simulations for fibers of various lengths and find
that the phenomenon persists regardless of the length of the fiber. Probably,
the most illuminating example is a \textquotedblleft fiber\textquotedblright%
\ of two coupled cells. Denoting the voltage at cell 1 as $V_{1}$ and that at
cell 2 as $V_{2}$, we can simulate the two coupled cells by integrating the
following equations \cite{KrassowskaEqns}
\begin{align}
\frac{dV_{1}}{dt}  &  =-\frac{1}{C_{m}}\left(  I_{\text{ion}}+I_{\text{ext}%
}\right)  +D\,\frac{\left(  V_{2}-V_{1}\right)  }{\Delta x^{2}}%
,\label{eqn:2cells_ode1}\\
\frac{dV_{2}}{dt}  &  =-\frac{1}{C_{m}}I_{\text{ion}}+D\,\frac{\left(
V_{1}-V_{2}\right)  }{\Delta x^{2}}. \label{eqn:2cells_ode2}%
\end{align}
Note that only cell 1 is paced in this case. Using various pacing protocols
and initial conditions, we find the two coupled cells can have spatially
desynchronized (Fig. \ref{fig:Dtype2} (a)) and synchronized (Fig.
\ref{fig:Dtype2} (b)) alternans. Note that in the desynchronized pattern, APD
in both cells exhibit a beat-to-beat variation of a few hundredth of ms, which
is impossible to observe in experiments.

\subsection{Influence of junctional Ca diffusion}

The original Shiferaw-Fox model does not include the diffusion of Ca between
neighboring cells \cite{CaCoupling}. Physiologically, there exists gap
junctional Ca diffusion although its magnitude is several orders lower than
voltage diffusion \cite{Bers2001,Allbritton1992Science} and thus is typically
neglected in cardiac modeling. One may wonder whether including junctional Ca
diffusion will affect the simulation results. To answer this question, we
introduce junctional Ca diffusion to the Shiferaw-Fox model by modifying the
equation governing [Ca$_{\text{in}}$] as follows \cite{KrassowskaEqns}:
\begin{equation}
\frac{\partial C_{i}}{\partial t}=D\,_{c}\frac{\partial}{\partial x}\left(
\frac{\partial C_{i}}{\partial x}-\frac{Z_{c}\,F}{R\,T}C_{i}\text{ }%
\frac{\partial v}{\partial x}\right)  -I_{c}, \label{eqn:Calcium}%
\end{equation}
where $C_{i}$ represents [Ca$_{\text{in}}$], $D_{c}=3\times10^{-9}$cm$^{2}/$ms
is the Ca diffusion coefficient \cite{Bers2001,Allbritton1992Science}, $Z_{c}$
is the valence of Ca, $F$ is the Faraday constant, $R$ is the gas constant,
$T=300$ K is the temperature, and $I_{c}$ represents the Ca$_{\text{in}}$
currents in the Shiferaw-Fox model. With this modification, we repeat the
numerical simulations and find there still exist multiple alternans solutions.

\begin{figure}[ptb]
\centering
\begin{tabular}{ccc}
\includegraphics[width=2.in]{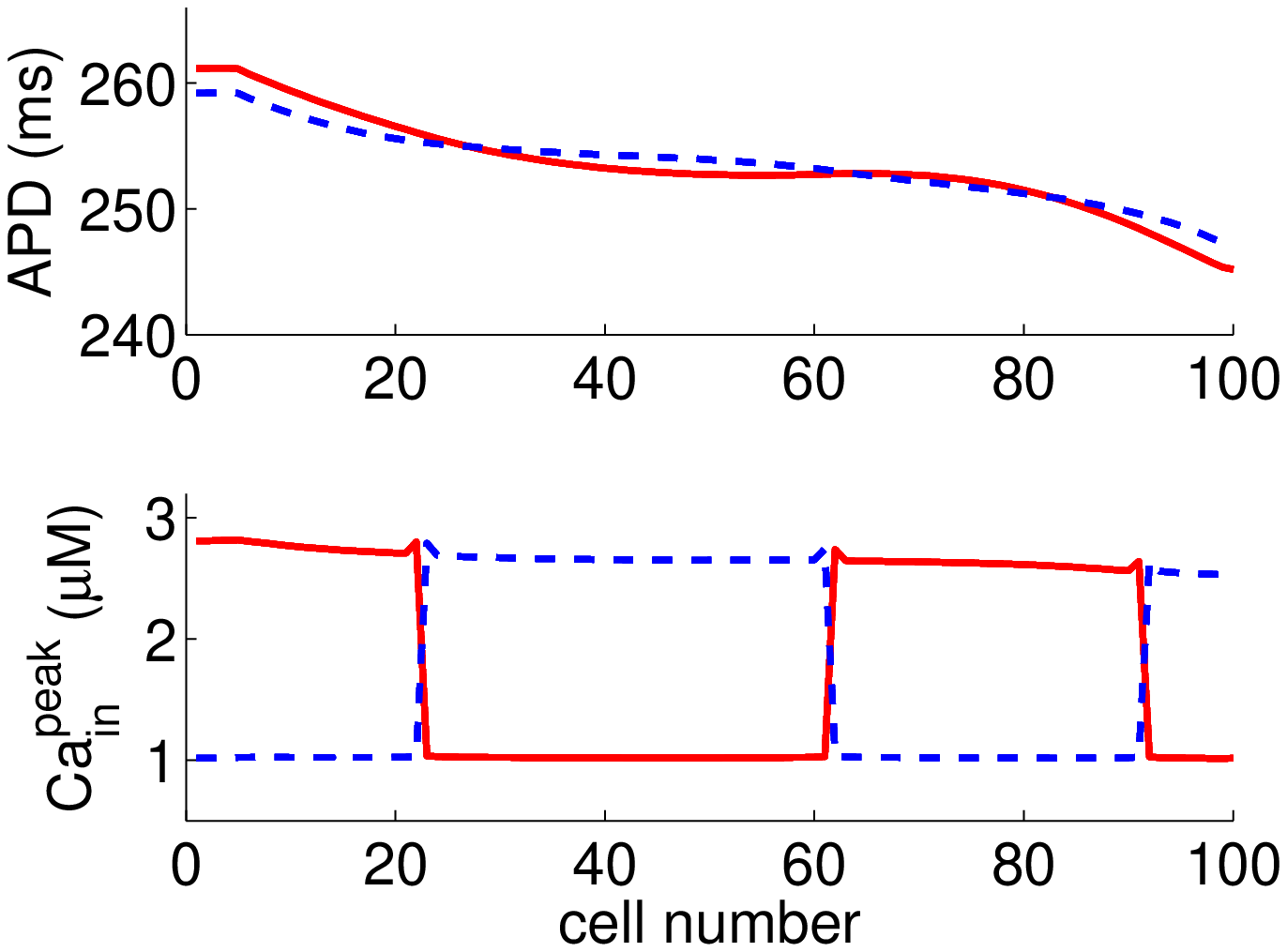} & %
\includegraphics[width=2.in]{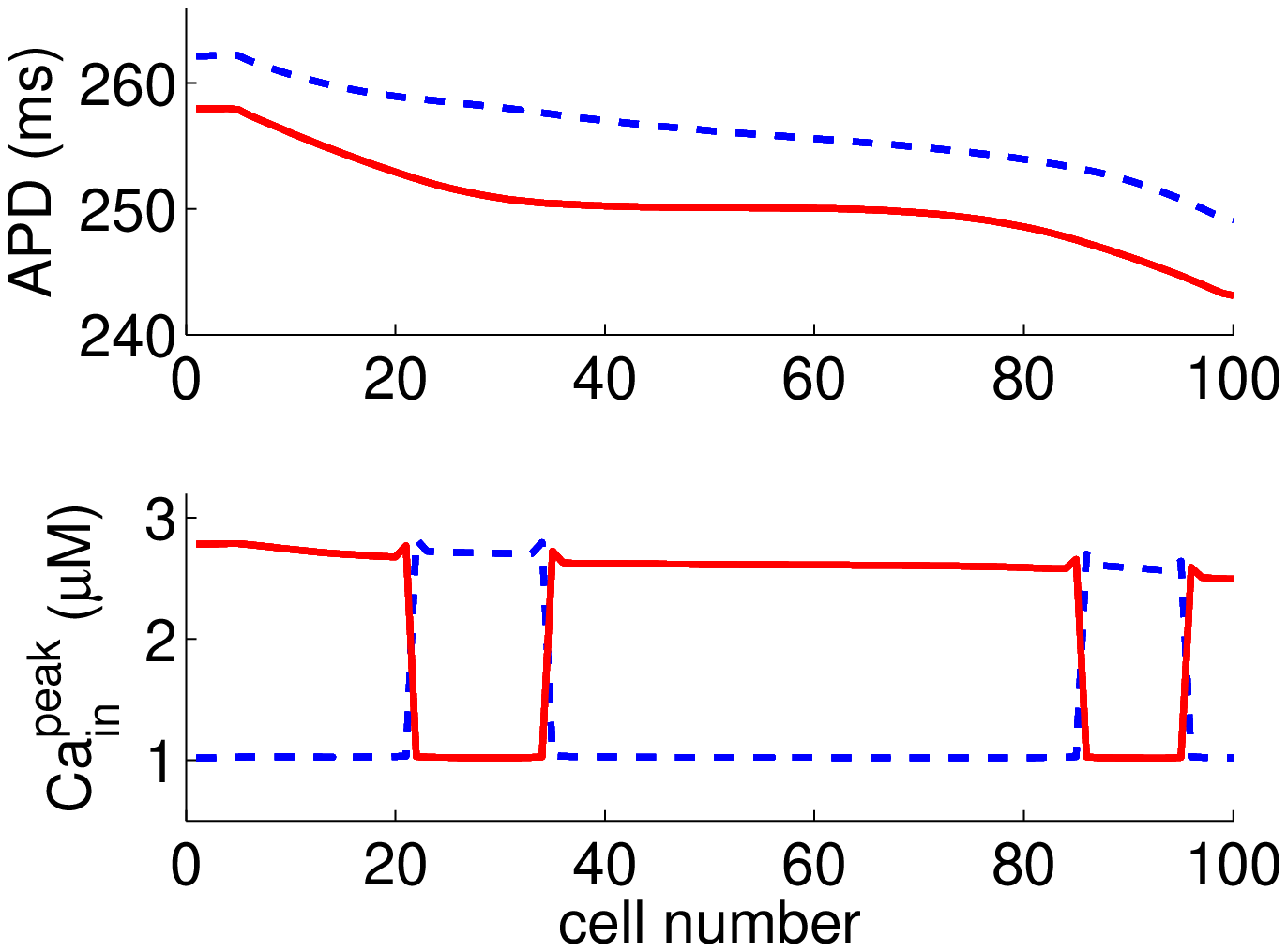} & %
\includegraphics[width=2.in]{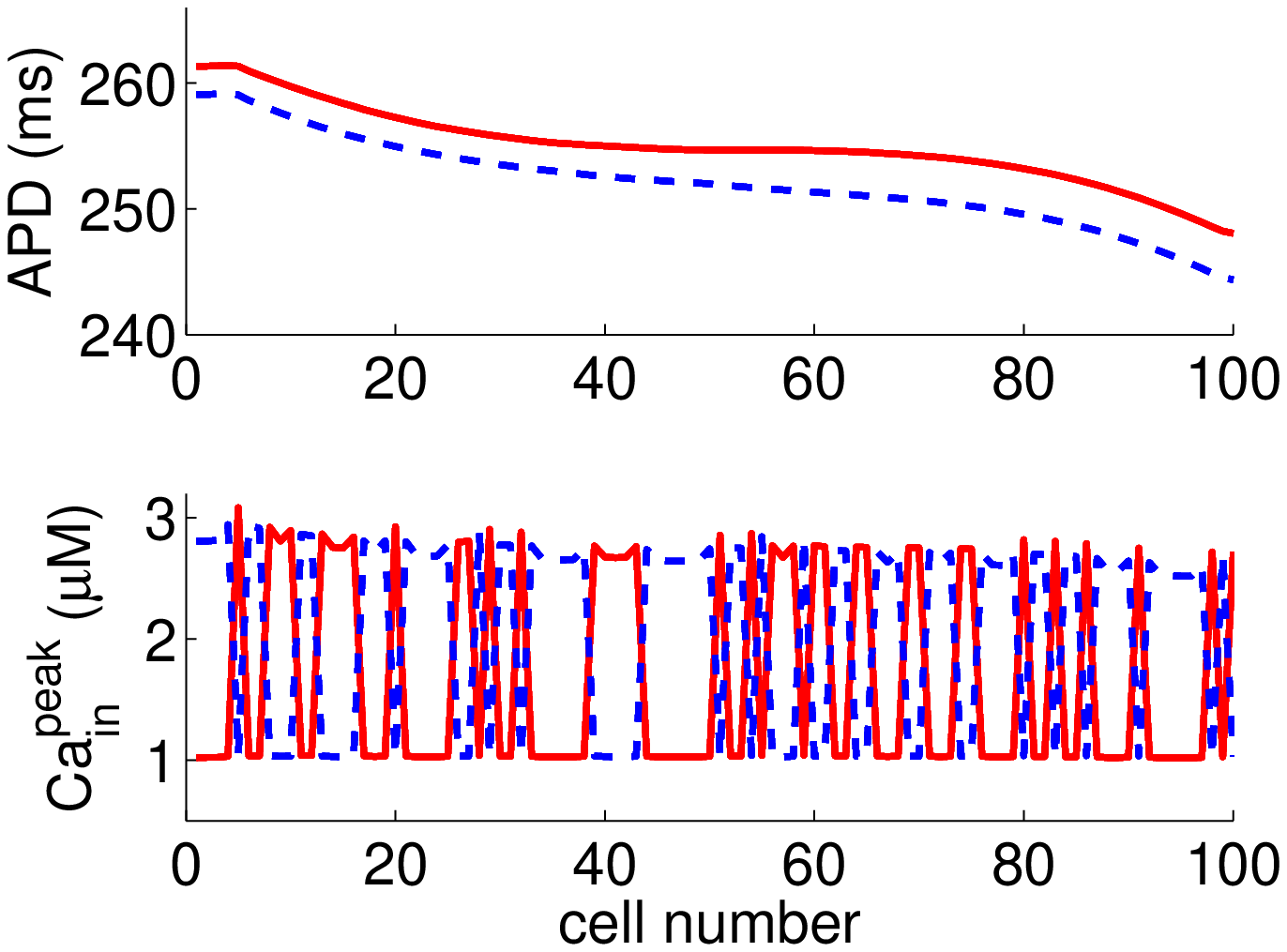} \\
(a) 499:-2:375 (ms) & (b) 500:-25:375 (ms) & (c) random distribution%
\end{tabular}%
\caption{Three selected alternans solutions for a homogeneous fiber
model, where junctional Ca diffusion is included. The cells have
negative Ca$\rightarrow V_{m}$ coupling and the pacing period is
$375$ ms. Panels (a) and (b) are obtained using downsweep protocols
and panel (c) is obtained by direct pacing with random initial
distribution of Ca$_{\text{in}}$. The last 10 beats of the
steady-state solutions are plotted, where odd beats are represented
by red solid lines and even beats by blue dashed lines.}
\label{fig:Dtype100_CaDiffusion}
\end{figure}

For example, Fig. \ref{fig:Dtype100_CaDiffusion} shows three selected
solutions for a homogeneous fiber of 100 cells paced at BCL=375 ms. Figures
\ref{fig:Dtype100_CaDiffusion} (a) and (b) are obtained using downsweep
protocols same as those in Fig. \ref{fig:Dtype100} (b) and (c), respectively.
Figure \ref{fig:Dtype100_CaDiffusion} (c) is obtained using a random initial
distribution in [Ca$_{\text{in}}$] (cf. Fig. \ref{fig:Dtype100} (f)).
Comparing results in Figs. \ref{fig:Dtype100} and
\ref{fig:Dtype100_CaDiffusion} shows that the coexistence of multiple
solutions is not influenced by junctional Ca diffusion.

\subsection{Influence of Ca$_{\text{in}}\rightarrow$V$_{\text{m}}$ coupling}

\begin{figure}[ptb]
\centering
\begin{tabular}{ccc}
\includegraphics[width=2.in]{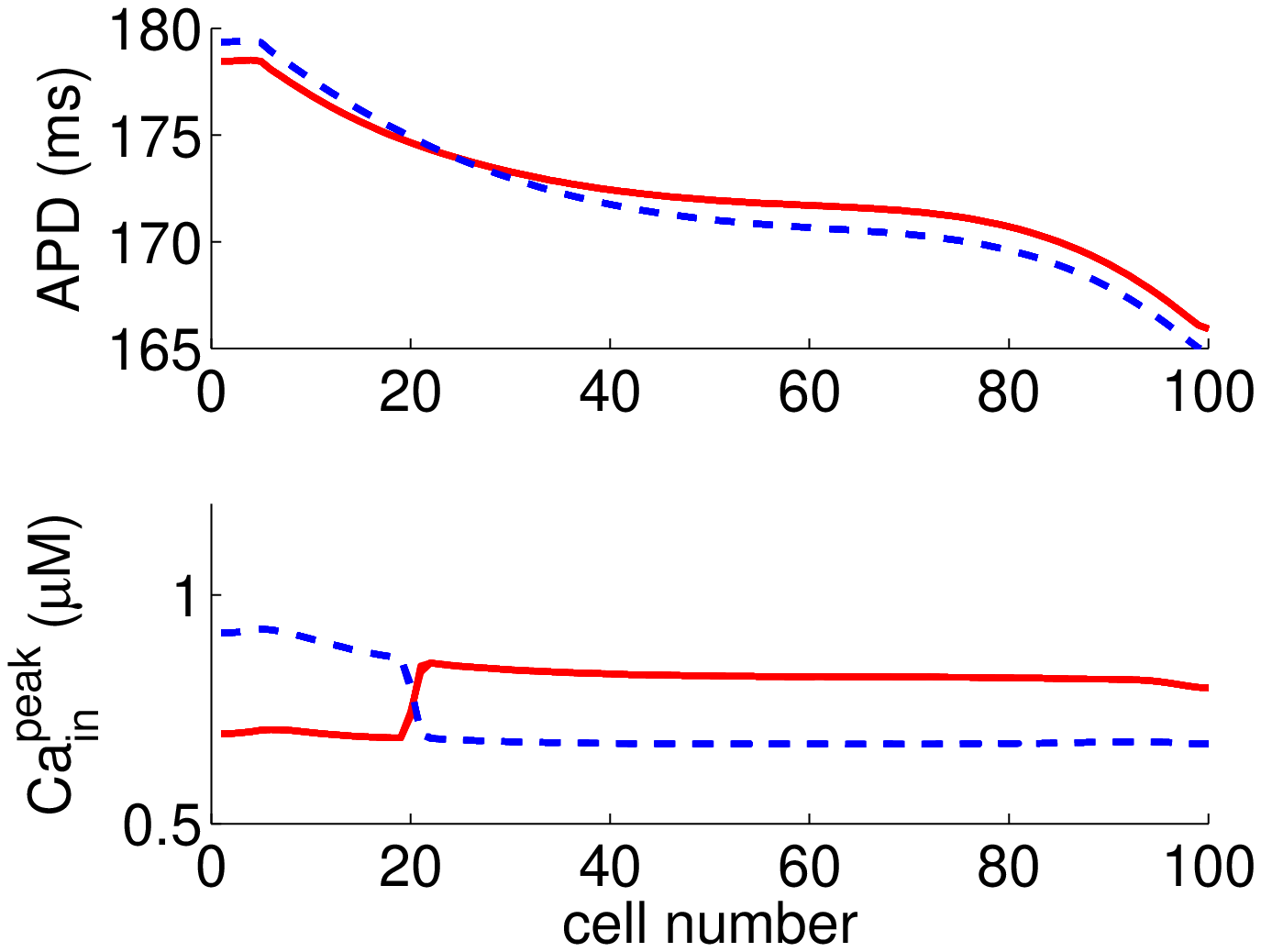} & %
\includegraphics[width=2.in]{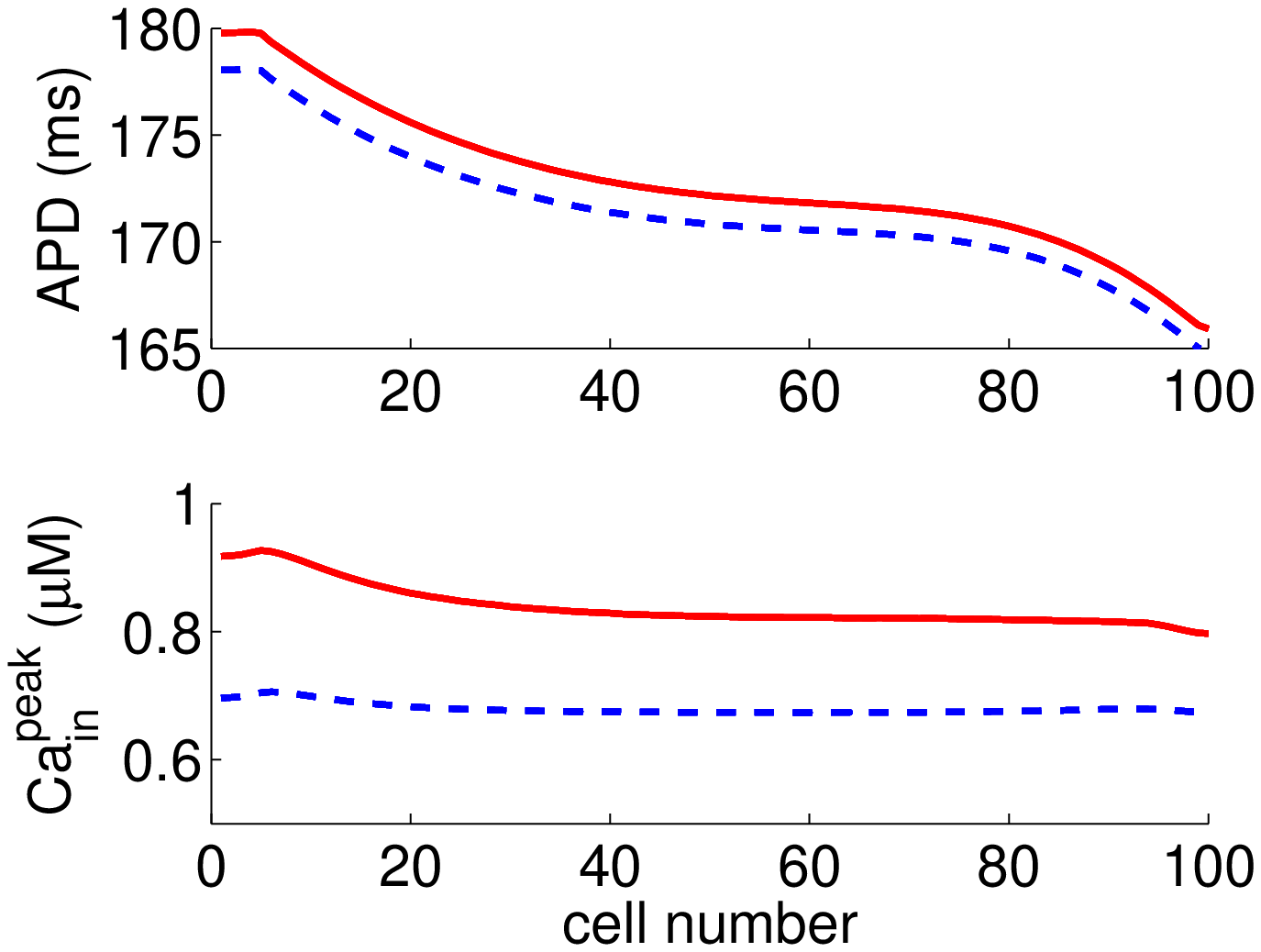} & %
\includegraphics[width=2.in]{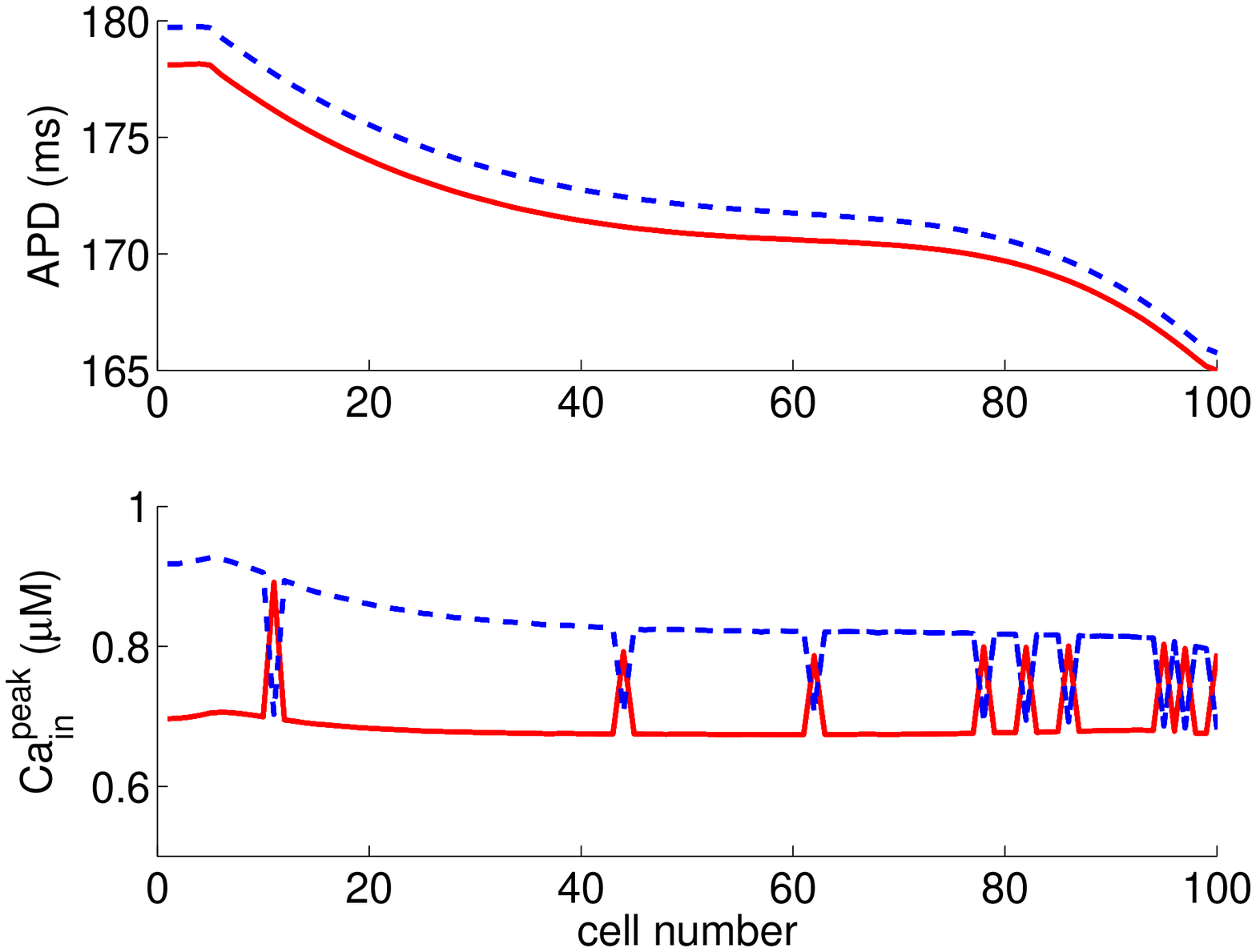} \\
(a) perturbed downsweep & (b) uniform distribution & (c) random distribution%
\end{tabular}%
\caption{Three selected alternans solution in a homogeneous fiber of
100 cells with positive Ca$\rightarrow V_{m}$ coupling. The pacing
period is $300$~ms in all panels. Panel (a) is obtained using a
perturbed downsweep protocol. Panels (b) and (c) are obtained by
direct pacing with with initial uniform distribution and random
distribution of Ca$_{\text{in}}$, respectively. (See text for
details.) The last 10 beats of the steady-state solutions are
plotted: odd beats are red solid and even beats are blue dashed.}
\label{fig:Ctype100}
\end{figure}

Simulations show that fibers with positive
Ca$_{\text{in}}\rightarrow $V$_{\text{m}}$ coupling also possess
multiple alternans solutions. For example, Fig. \ref{fig:Ctype100}
shows 3 selected solutions for a homogeneous fiber of 100 cells with
positive Ca$_{\text{in}}\rightarrow$V$_{\text{m}}$ coupling, paced
at BCL=300 ms. Figure \ref{fig:Ctype100} (a) is obtained via a
perturbed downsweep protocol, where the step size is $\Delta$B=25 ms
and a long and a short perturbations of $\pm20$ ms are applied at
each BCL, see section 2 for details of the perturbed downsweep
protocol. Figures \ref{fig:Ctype100} (b) and (c) are obtained by
directly pacing the cell at BCL=300 ms using a uniform and a random
initial distributions in [Ca$_{\text{in}}$], respectively.

\begin{figure}[tbh]
\centering
\begin{tabular}
[c]{cc}%
\includegraphics[width=2.in]{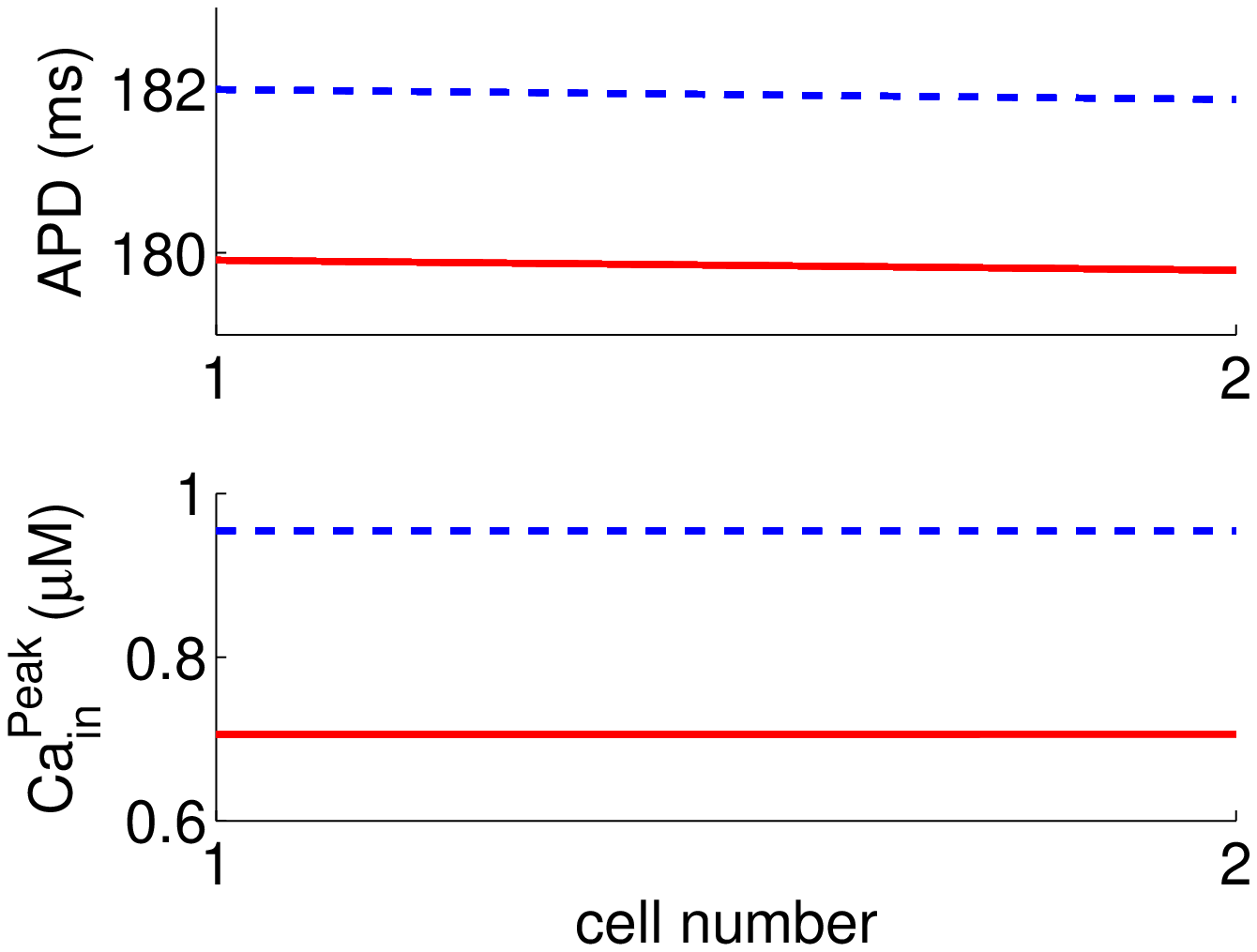} &
\includegraphics[width=2.in]{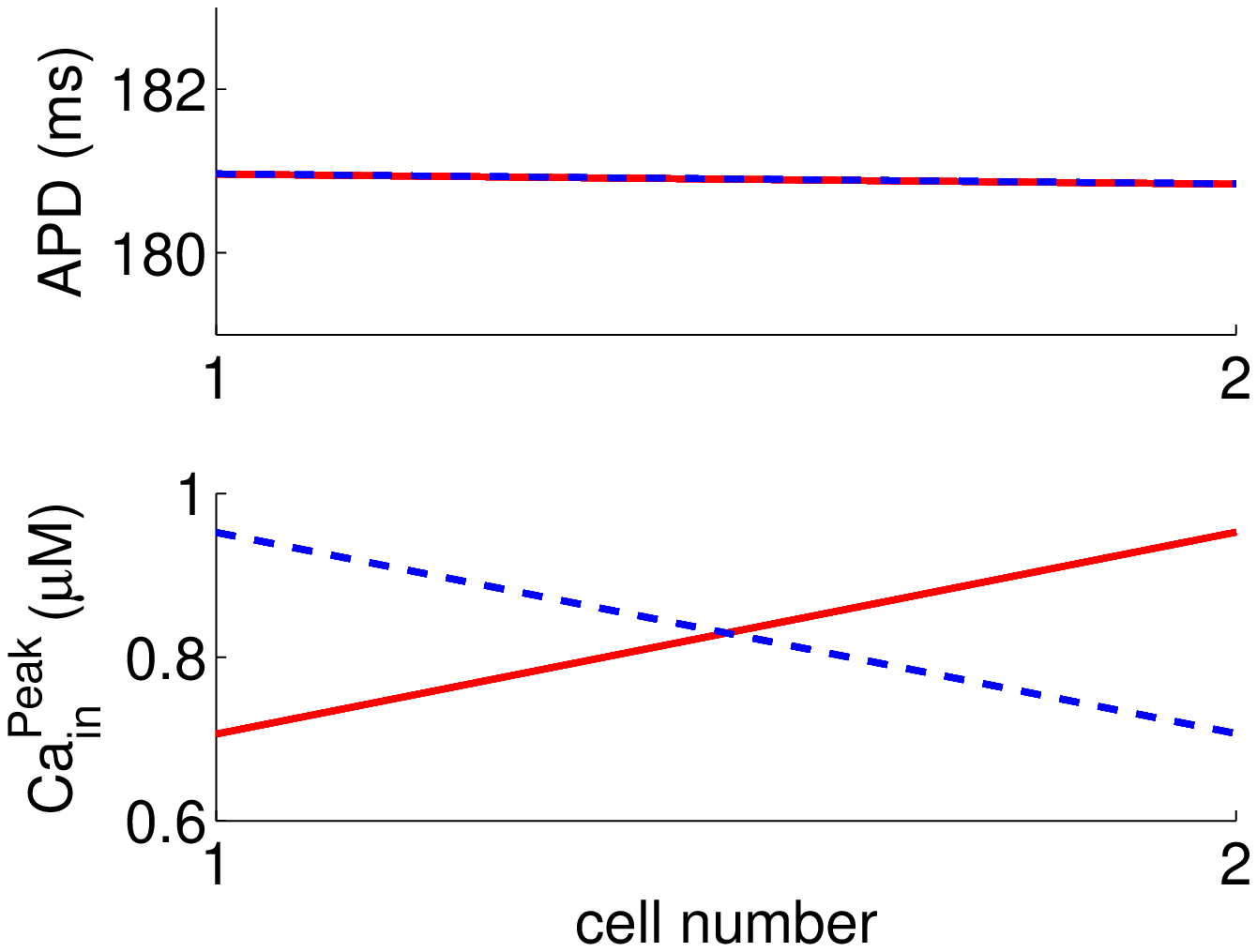}\\
(a) & (b)
\end{tabular}
\caption{(Color online) A \textquotedblleft fiber\textquotedblright\ of two
cells with positive Ca$\rightarrow V_{m}$ coupling can have both spatially
synchronized and desynchronized alternans solutions. The pacing period is
BCL$=300$~ms. The last 10 beats of the steady-state solutions are plotted: odd
beats are red solid and even beats are blue dashed.}%
\label{fig:Ctype2}%
\end{figure}

We simulate two coupled cells with positive Ca$_{\text{in}}$ $\rightarrow$
V$_{\text{m}}$ coupling and find that alternans can be both spatially
synchronized and spatially desynchronized, see Fig. \ref{fig:Ctype2}. In Figs.
\ref{fig:Dtype2} and \ref{fig:Ctype2}, values of APD are almost identical in
the two cells because conduction time across a cell's length is negligible.
Thus, these examples indicate that the coexistence of multiple solutions does
not depend on steep CV\ restitution. We note that, using a downsweep protocol
with step size of 2 ms, Sato et al. \cite{Sato2006CircRes} also studied
alternans in two coupled cells. They observed the desynchronized solution for
negative Ca$_{\text{in}}\rightarrow$V$_{\text{m}}$ coupling and the
synchronized solution for positive Ca$_{\text{in}}\rightarrow$V$_{\text{m}}$
coupling. Using numerical simulation, Sato et al. showed that negative
Ca$_{\text{in}}\rightarrow$V$_{\text{m}}$ coupling tends to desynchronize two
coupled cells whereas positive Ca$_{\text{in}}\rightarrow$V$_{\text{m}}$
coupling tends to synchronize them. Thus, we hypothesize that the synchronized
solution for negative Ca$_{\text{in}}\rightarrow$V$_{\text{m}}$ coupling and
the desynchronized one for positive coupling are induced by electrotonic
coupling, which will be verified in the next section.

\section{Origin of the Indeterminacy}

\subsection{Numerically tracing an alternans solution}

The coexistence of multiple alternans solutions in a fiber is surprising
because the underlying cell's model possesses a supercritical period-doubling
bifurcation, describing a transition from a unique 1:1 solution to a unique
alternans solution (Fig. \ref{fig:ode_bif}). The transition to alternans in a
fiber appears to be more complicated. First, when paced at sufficiently large
values of BCL, a fiber has a single 1:1 solution, regardless of the initial
condition. However, as shown in the previous section, the fiber can develop
multiple patterns of alternans. Therefore, it is interesting to ask how
different alternans patterns are related to one another and how they are
connected to the 1:1 solution. To address these questions, we utilize the
\emph{upsweep protocol} to trace the \textquotedblleft
origins\textquotedblright\ of different alternans solutions. Starting at
BCL=$375$ ms from a pattern in Fig. \ref{fig:Dtype100}, we increase BCL by 1
ms every 200 beats until the alternans solution either changes to a different
alternans pattern or becomes a 1:1 pattern. Then, we mark that value of BCL as
the \textquotedblleft starting\textquotedblright\ point for the studied
alternans pattern. Under this protocol, each solution in Fig.
\ref{fig:Dtype100} transforms into another alternans solution at a starting
BCL shown in Table \ref{tab:Dtype100_origins}. Specifically, pattern (a)
\textquotedblleft starts\textquotedblright\ at BCL=$416$ ms and patterns
(b)-(f) start at BCL values between $402$ and $405$ ms. Further numerical
simulations show that the fiber first undergoes alternans at BCL=$425$ ms.
This is counterintuitive because, in the underlying cell's model, the onset of
alternans occurs at $401$ ms (cf. Fig. \ref{fig:ode_bif}). Since alternans
solutions in the fiber can occur earlier than alternans in the single-cell
model, we hypothesize that these solutions are induced from the interaction
between cellular dynamics and electrotonic coupling (\emph{induced
alternans}). We further hypothesize that alternans may also result from the
intrinsic instability mechanism that leads to alternans in the underlying
cellular model (\emph{intrinsic alternans}). Noticing that the spatially
concordant alternans in pattern (d) starts at BCL=$402$ ms, near the onset of
alternans in the single-cell model, we hypothesize the concordant alternans is
a result of intrinsic alternans. \begin{table}[tbh]
\caption{Origins of the alternans patterns in Fig. \ref{fig:Dtype100}.}%
\label{tab:Dtype100_origins}
\centering%
\begin{tabular}
[c]{|c||c|c|c|c|c|c|}\hline
pattern & (a) & (b) & (c) & (d) & (e) & (f)\\\hline\hline
starting BCL (ms) & 416 & 405 & 404 & 402 & 403 & 404\\\hline
\end{tabular}
\end{table}

Using upsweep protocols, we also trace the origins of the alternans in Fig.
\ref{fig:Dtype2} for the two coupled cells. Numerical simulations show that
the desynchronized solution in Fig. \ref{fig:Dtype2} (a) stops at BCL=405 ms
and the synchronized solution in Fig. \ref{fig:Dtype2} (b) stops at BCL=401
ms. Moreover, we also find that different alternans solutions for fibers with
positive Ca$_{\text{in}}\rightarrow$V$_{\text{m}}$ coupling stop at different
values of BCL. Thus, the bifurcation structure for a fiber is much more
complicated than that for a single cell. Moreover, the numerical results
suggest that the multiple alternans patterns born from different bifurcations
are due to the interaction between electrotonic coupling and instability in
calcium cycling.

\subsection{Bifurcation Mechanism}

Shiferaw et al. showed that the bidirectional V$_{\text{m}}$/Ca coupling in
the Shiferaw-Fox can be captured by a 2-d mapping model \cite{Shiferaw2005PRE}%
. The work of Shiferaw et al. provides a theoretical framework for
understanding various experimental observations including electromechanically
in-phase/out-of-phase alternans and quasi-periodic oscillations of voltage and
calcium. Here, we follow their approach and take a mapping model in the
following form \cite{SatoCommunication},
\begin{align}
A_{n+1}  &  =f\left(  D_{n}\right)  +\gamma\,\left(  C_{n+1}-C_{\text{crit}%
}\right)  ,\label{eqn:bcmap_A}\\
C_{n+1}  &  =\mu\,D_{n}-g\left(  C_{n}\right)  , \label{eqn:bcmap_C}%
\end{align}
where $A_{n}$, $D_{n}$, and $C_{n}$ represent the APD, DI, and peak
[Ca$_{\text{in}}$] concentration of $n$th beat, respectively. By definition,
it follows that $D_{n}=$BCL$-A_{n}$, see Fig. \ref{fig:action_potential}.
Function $f$ stands for the APD resitution. The second term in Eq.
(\ref{eqn:bcmap_A}) accounts for the influence of [Ca$_{\text{in}}$] on APD.
Negative Ca$_{\text{in}}\rightarrow$V$_{\text{m}}$ coupling corresponds to
$\gamma<0$ and positive coupling to $\gamma>0$. Due to graded release, $\mu$
has to be positive \cite{Sato2006CircRes}. Function $g$ determines the
relation between Ca concentrations in two consecutive beats.

Using map (\ref{eqn:bcmap_A}, \ref{eqn:bcmap_C}), we will show how the
interaction between electrotonic coupling and instability in calcium cycling
leads to multiple alternans patterns in fibers. Since the analysis below does
not depend on the forms of $f$ and $g$, we do not specify their forms here.
Instead we present numerical examples based on concrete forms of $f$ and $g$
in the Appendix. The numerical examples there show that fiber models based on
the map (\ref{eqn:bcmap_A}, \ref{eqn:bcmap_C}) are able to reproduce the
phenomena observed in simulations of the Shiferaw-Fox model.

\subsubsection{Bifurcation for single cells}

Using a mapping model similar to (\ref{eqn:bcmap_A},\ref{eqn:bcmap_C}),
Shiferaw et al. \cite{Shiferaw2005PRE} carried out a bifurcation analysis to
show how different alternans solutions in single cells arise as a result of
V$_{\text{m}}$/Ca coupling. Here, we briefly review that analysis, which
serves as a starting point to understand bifurcation for fibers. To this end,
consider a paced cell described by map (\ref{eqn:bcmap_A},\ref{eqn:bcmap_C}).
A 1:1 solution is a fixed point of the map and its stability is determined by
the Jacobian matrix:%
\begin{equation}
J=\left(
\begin{array}
[c]{cc}%
-f^{\prime}-\gamma\,\mu\,\,\, & \,\,\,-\gamma\,g^{\prime}\\
-\mu\,\,\, & \,\,\,-g^{\prime}%
\end{array}
\right)  ,
\end{equation}
where $f^{\prime}$ measures the slope of the APD restitution and $g^{\prime}$
measures the slope of the [Ca$_{\text{in}}$] relation, and all derivatives are
evaluated at the fixed point.

The strength of V$_{\text{m}}$/Ca coupling varies among species
\cite{Shiferaw2005PRE}. In this paper, we assume a weak coupling between
V$_{\text{m}}$ dynamics and Ca cycling, which allows easy analysis and yields
insight on the instability mechanisms in cardiac cells as well as in fibers.
To this end, we assume $\gamma\,\mu<<f^{\prime}$, $g^{\prime}$ in the
following. Using a perturbation technique \cite{Nayfeh1973Perturbation}, we
find the eigenvalues of $J$ to first order in $\gamma\,\mu$ are
\begin{equation}
-f^{\prime}\,\left(  1+\frac{\gamma\,\mu}{f^{\prime}-g^{\prime}}\right)
\,\,\text{and\thinspace\thinspace}-g^{\prime}\,\left(  1+\frac{\,\gamma\,\mu
}{g^{\prime}-f^{\prime}}\right)  . \label{eqn:eigvals_singlecell}%
\end{equation}
Since $\gamma\,\mu$ is small, one can see that a period-doubling bifurcation
occurs if one of the slopes, $f^{\prime}$ or $g^{\prime}$, becomes sufficient
large (compared to $1$). Thus, if the bifurcation is caused by increasing in
$f^{\prime}$, we say the alternans is APD driven. On the other hand, it is
called a Ca-driven alternans if the bifurcation is due to increasing in
$g^{\prime}$. \ For simplicity, if alternans is APD driven, we assume
$g^{\prime}$ remains small and will not induce dynamic instability for all
physiologically interesting parameter regimes; and vice versa.

\textbf{APD-driven alternans:} In case of APD-driven alternans, it follows
that $f^{\prime}>g^{\prime}$. To first order in $\gamma\,\mu$, the unstable
eigenvector is
\begin{equation}
\left(
\begin{array}
[c]{cc}%
1,\,\,\, & \frac{1}{f^{\prime}-g^{\prime}}-\frac{\gamma\,\mu\,f^{\prime}%
}{\left(  f^{\prime}-g^{\prime}\right)  ^{3}}%
\end{array}
\right)  ^{T},
\end{equation}
which is in phase in APD and [Ca$_{\text{in}}$]. Therefore, APD-driven
alternans is electromechanically in phase.

\textbf{Calcium-driven alternans: }In case of Ca-driven alternans, it follows
that $g^{\prime}>f^{\prime}$. To first order in $\gamma\,\mu$, the unstable
eigenvector is
\begin{equation}
\left(
\begin{array}
[c]{cc}%
\frac{\gamma\,\mu\,g^{\prime}}{g^{\prime}-f^{\prime}},\,\,\, & 1
\end{array}
\right)  ^{T}.
\end{equation}
For positive Ca$_{\text{in}}\rightarrow$V$_{\text{m}}$ coupling, it follows
that $\gamma>0$ and thus alternans is electromechanically in phase. On the
other hand, for negative Ca$_{\text{in}}\rightarrow$V$_{\text{m}}$ coupling,
it follows that $\gamma<0$ and thus alternans is electromechanically out of phase.

\subsubsection{Bifurcation for two coupled cells}

To understand the bifurcation mechanism for fibers, we start with the extreme
case of two coupled cells. Denoting the locations of the cells by $x_{1}$ and
$x_{2}$, we represent the APD, DI, and [Ca$_{\text{in}}$] of cell $i$ for the
$n$th beat as $A_{n}\left(  x_{i}\right)  $, $D_{n}\left(  x_{i}\right)  $,
and $C_{n}\left(  x_{i}\right)  $, respectively. Following Fox et al.
\cite{Fox2002CircRes}, we account for electrotonic coupling between the cells
using weighted averaging:%
\begin{align}
A_{n+1}\left(  x_{1}\right)   &  =\rho\left(  0\right)  \,F\left(
x_{1}\right)  +\rho\left(  1\right)  \,F\left(  x_{2}\right)
\label{eqn:2cells_map1}\\
A_{n+1}\left(  x_{2}\right)   &  =\rho\left(  1\right)  \,F\left(
x_{1}\right)  +\rho\left(  0\right)  \,F\left(  x_{2}\right) \\
C_{n+1}\left(  x_{1}\right)   &  =\mu\,D_{n}\left(  x_{1}\right)  -g\left(
C_{n}\left(  x_{1}\right)  \right) \\
C_{n+1}\left(  x_{2}\right)   &  =\mu\,D_{n}\left(  x_{2}\right)  -g\left(
C_{n}\left(  x_{2}\right)  \right)  , \label{eqn:2cells_map4}%
\end{align}
where
\begin{equation}
F\left(  x_{i}\right)  =f\left(  D_{n}\left(  x_{i}\right)  \right)
+\gamma\,\left(  C_{n+1}\left(  x_{i}\right)  -C_{\text{crit}}\right)  ,
\end{equation}
and $\rho\left(  0\right)  $ and $\rho\left(  1\right)  $ are the weighting
functions. Here, we have neglected the effect of dispersion due to conduction
velocity restitution since the \textquotedblleft fiber\textquotedblright\ is
short and thus the influence of conduction restitution to APD is
insignificant. On the time scale of one APD, V$_{m}$ diffuses on a spatial
scale of a few tens of cells \cite{Echebarria2002PRL}. Thus, the coupling in
APD between two neighboring cells is strong. As a result, $\rho\left(
0\right)  \approx\rho\left(  1\right)  $ and $\rho\left(  0\right)
+\rho\left(  1\right)  =1$. Therefore, we let $\rho\left(  0\right)
=1/2+\varepsilon$ and $\rho\left(  1\right)  =1/2-\varepsilon$, where
$0<\varepsilon<<1$. For a 1:1 solution, the two cells have the same values of
APD and [Ca$_{\text{in}}$]. Thus, the Jacobian matrix can be written as%
\begin{equation}
J=\left(
\begin{array}
[c]{cccc}%
\rho\left(  0\right)  \,\left(  -f^{\prime}-\gamma\,\mu\right)  &
\,\,\,\rho\left(  1\right)  \,\left(  -f^{\prime}-\gamma\,\mu\right)  &
\,\,\,-\rho\left(  0\right)  \,\gamma\,g^{\prime} & \,\,\,-\rho\left(
1\right)  \,\gamma\,g^{\prime}\\
\rho\left(  1\right)  \,\left(  -f^{\prime}-\gamma\,\mu\right)  &
\,\,\,\rho\left(  0\right)  \,\left(  -f^{\prime}-\gamma\,\mu\right)  &
\,\,\,-\rho\left(  1\right)  \,\gamma\,g^{\prime} & \,\,\,-\rho\left(
0\right)  \,\gamma\,g^{\prime}\\
-\mu & \,\,\,0 & \,\,\,-g^{\prime} & \,\,\,0\\
0 & \,\,\,-\mu & \,\,\,0 & \,\,\,-g^{\prime}%
\end{array}
\right)  , \label{eqn:jacobi_2cells}%
\end{equation}
where all derivatives are evaluated at the 1:1 solution.

Using perturbation theory \cite{Nayfeh1973Perturbation}, we find the
eigenvalues to first order in $\gamma\,\mu$ and $\varepsilon$ are%
\begin{equation}
-f^{\prime}\,\left(  1+\frac{\gamma\,\mu}{f^{\prime}-g^{\prime}}\right)
,\,\,-g^{\prime}\,\left(  1+\frac{\,\gamma\,\mu}{g^{\prime}-f^{\prime}%
}\right)  ,\,\,-2\,\varepsilon\,f^{\prime},\,\text{and}\,-g^{\prime
}-2\,\varepsilon\,\gamma\,\mu. \label{eqn:eigvals_2cells}%
\end{equation}
Note that the first two eigenvalues are the same as those for the case of
single cells (cf. Eq. \ref{eqn:eigvals_singlecell}). Thus, they are due to
intrinsic membrane dynamics. The last two eigenvalues are induced by the
electrotonic coupling as manifested by the small coupling parameter
$\varepsilon$.

\textbf{APD-driven alternans: }For APD-driven alternans, the intrinsic
bifurcation is the only mechanism to produce alternans. The corresponding
eigenvector to first order in $\gamma\,\mu$ and $\varepsilon$ is
\begin{equation}
\left(
\begin{array}
[c]{cccc}%
1,\,\,\, & 1,\,\,\, & \frac{1}{f^{\prime}-g^{\prime}}-\frac{\gamma
\,\mu\,f^{\prime}}{\left(  f^{\prime}-g^{\prime}\right)  ^{3}},\,\,\, &
\frac{1}{f^{\prime}-g^{\prime}}-\frac{\gamma\,\mu\,f^{\prime}}{\left(
f^{\prime}-g^{\prime}\right)  ^{3}}%
\end{array}
\right)  ^{T}.
\end{equation}
Thus, APD-driven alternans tends to be electro-mechanically in phase and
spatially synchronized.

\textbf{Calcium-driven alternans: }For Ca-driven alternans, bifurcation can
happen intrinsically when $g^{\prime}+\,\gamma\,\mu\,g^{\prime}/\left(
g^{\prime}-f^{\prime}\right)  =1$ (\emph{intrinsic bifurcation}). Bifurcation
can also be induced by electrotonic coupling when $g^{\prime}+2\,\varepsilon
\,\gamma\,\mu=1$ (\emph{induced bifurcation}). To first order in $\gamma\,\mu$
and $\varepsilon$, the eigenvector corresponding to the intrinsic bifurcation
is%
\begin{equation}
\left(
\begin{array}
[c]{cccc}%
\frac{\gamma\,\mu\,g^{\prime}}{g^{\prime}-f^{\prime}},\,\,\, & \frac
{\gamma\,\mu\,g^{\prime}}{g^{\prime}-f^{\prime}},\,\,\, & 1,\,\,\, & 1
\end{array}
\right)  ^{T}. \label{eqn:2cells_eiginphase}%
\end{equation}
To first order in $\gamma\,\mu$ and $\varepsilon$, the eigenvector
corresponding to the induced bifurcation is
\begin{equation}
\left(
\begin{array}
[c]{cccc}%
2\,\varepsilon\,\gamma\,\mu,\,\,\, & -2\,\varepsilon\,\gamma\,\mu,\,\,\, &
1,\,\,\, & -1
\end{array}
\right)  ^{T}. \label{eqn:2cells_eigoutphase}%
\end{equation}
Therefore, while the intrinsic bifurcation gives birth to a spatially
synchronized solution, the induced bifurcation leads to a spatially
desynchronized pattern. And the electromechanical phase is determined by the
sign of $\gamma$.

Subtracting the intrinsic eigenvalue from the induced one yields $\gamma
\,\mu\,\left(  g^{\prime}\,\left(  g^{\prime}-f^{\prime}\right)
-2\varepsilon\right)  $, which has the same sign as $\gamma$ since $g^{\prime
}>f^{\prime}$ (Ca-driven alternans) and $\mu>0$ (graded release). If
$\gamma<0$, the induced eigenvalue is more negative; therefore, the spatially
desynchronized pattern, born from the induced bifurcation, occurs before the
spatially synchronized pattern, born from the intrinsic bifurcation. On the
other hand, if $\gamma>0$, the intrinsic eigenvalue is more negative;
therefore, the spatially desynchronized pattern occurs after the spatially
synchronized pattern. The analysis is in agreement with numerical simulations
of the Shiferaw-Fox model using upsweep protocols. Therefore, when BCL is
continuously decreased from a 1:1 solution, it is more likely to observe the
spatially synchronized solution for the positive Ca$_{\text{in}}\rightarrow
$V$_{\text{m}}$ coupling case whereas it is more likely to observe spatially
desynchronized pattern in the negative Ca$_{\text{in}}\rightarrow$%
V$_{\text{m}}$ coupling case. This is probably why Sato et al. observed
different solution patterns for negative and positive Ca$_{\text{in}%
}\rightarrow$V$_{\text{m}}$ couplings in their downsweep simulations.

The conclusion based on two coupled cells may be extended to short fibers;
however, we expect long fibers have more complicated phenomena than presented
here. In particular, because the dispersion effect may play an important role
in long fibers as shown by Echebarria and Karma
\cite{Echebarria2002PRL,Echebarria2006} and by Dai and Schaeffer
\cite{Dai2008}. Nevertheless, the simple case sheds light on the effect of
electrotonic coupling, the difference between APD and Ca-driven alternans, and
the difference between negative and positive Ca$_{\text{in}}\rightarrow
$V$_{\text{m}}$ couplings.

\subsubsection{Understanding bifurcation for fibers}

For Ca-driven alternans in fibers, we start with the extreme case of $\mu=0$.
As can be seen from the cellular model (\ref{eqn:bcmap_A},\ref{eqn:bcmap_C}),
in this case, Ca$_{\text{in}}$ dynamics becomes independent of V$_{\text{m}}$
dynamics. Because Ca$_{\text{in}}$ dynamics is not influenced by electrotonic
coupling, when alternans develops in a fiber, cells can arbitrarily choose
their phases in [Ca$_{\text{in}}$]. Depending on the initial conditions, Ca
alternans on a fiber can take different spatial patterns. When $\mu$ is small
but nonzero, [Ca$_{\text{in}}$] in different cells are weakly coupled due to
feedback of V$_{\text{m}}$ dynamics and electrotonic coupling. If this weak
coupling does not suppress the coexistence of multiple solutions, the
alternans pattern will become sensitive to initial conditions and pacing
protocols. Under certain pacing protocols, Ca$_{\text{in}}$ alternans on a
fiber may possess a complex pattern with multiple phase reversals, as shown in
Figs. \ref{fig:Dtype100} (f), \ref{fig:Dtype100_CaDiffusion} (c), and
\ref{fig:Ctype100} (c). In these examples, the spatial patterns of APD are
much less complex, which is because APD on the fiber is an averaging effect
over many cells due to the fast diffusion of V$_{\text{m}}$
\cite{Echebarria2002PRL,Echebarria2006}. As a result of electrotonic coupling,
different spatial patterns of Ca$_{\text{in}}$ alternans may correspond to
similar spatial patterns of APD. Moreover, as shown in simulations of the
Shiferaw-Fox model (see Figs. \ref{fig:Dtype100},
\ref{fig:Dtype100_CaDiffusion}, and \ref{fig:Ctype100}) and of the mapping
model (see Fig. \ref{fig:negmap_100cells} in Appendix), spatial dyssynchrony
in Ca$_{\text{in}}$ alternans may induce spatially discordant APD alternans,
which verifies the hypothesis of Aistrup et al. \cite{Aistrup2006CircRes}.

\section{Summary and Discussion}

Using numerical simulation and theoretical analysis, we have investigated
spatiotemporal patterns of calcium-driven alternans. The main finding is that
although alternans of an isolated cell is solely determined by the pacing
period, the solution in a fiber is sensitive to pacing protocols and initial
conditions. To the author's knowledge, this is the first report on the
coexistence of multiple alternans solutions for cardiac fibers. We have
further verified that the coexistence of multiple alternans solutions is
independent of the length of the fiber, junctional Ca diffusion, and the type
of Ca$_{\text{in}}\rightarrow$V$_{\text{m}}$ coupling. Since multiple
solutions also exist for fibers of as few as a couple of cells, the phenomenon
does not require steep conduction velocity restitution. Another interesting
observation is that complex patterns of Ca$_{\text{in}}$ alternans with
multiple phase reversals in neighboring cells may arise in a homogeneous fiber
with both negative and positive Ca$_{\text{in}}\rightarrow$V$_{\text{m}}$
couplings. The simulation results also verify the hypothesis of Aistrup et al.
that spatially desynchronized Ca signaling may lead to spatially discordant
APD alternans \cite{Aistrup2006CircRes}.

We have also explored the bifurcation mechanism for the coexistence of
multiple solutions. Numerical simulations and symbolic bifurcation analyses
trace the onset of multiple alternans patterns to a number of bifurcations
induced by the interaction between electrotonic coupling and an instability in
calcium cycling. The bifurcation here bears similiarity to that of a ring
model. For example, Courtmanche et al. \cite{Courtemanche1993PRL} used the
Beeler-Reuter model \cite{BeelerJPhysio1977} to show that, when the length of
the ring is reduced, there is an infinite-dimensional Hopf bifurcation, which
gives birth to an infinite number of quasi-periodic solutions. Since the
Beeler-Reuter model does not account for intracellular calcium cycling, it
will be interesting to investigate how calcium-induced alternans propagates in
a ring.

Simulations in this paper lead to another interesting observation---the onset
of altrenans in a fiber does not occur at the same value of BCL as that in the
underlying single-cell model, a phenomenon against the common wisdom. This
counter-intuitive observation indicates that one can not directly relate
behavior of fibers to that of single cell models. Another intriguing example
about the differences between alternans in fibers and alternans in single
cells is presented in the work of Cherry and Fenton \cite{Cherry2007AmJPhy},
who analyzed two models of canine ventricular myocytes: the
Fox-MchHarg-Gilmour (FMG) model \cite{Fox2002AmJP} and the Hund-Rudy (HR)
model \cite{Hund2004Cir}. Simulations of both models show that the bifurcation
structures for fibers are different than those for the corresponding single
cells (see Fig. 5 in \cite{Cherry2007AmJPhy}). Most interestingly, the HR 0d
model (a single cell) shows alternans for BCL between 180 and 230 ms whereas
the HR 1d model (a 1.25 cm-long fiber) does not undergo alternans for all
values of BCL studied. Examples in \cite{Cherry2007AmJPhy} as well as examples
in the current paper show that stability of a fiber may not be inferred from
stability of the single cell model and vice versa.

Results in this paper are based on numerical simulations and hypothesized
mapping models. Here, we describe potential experiments to verify the
numerical observations. Probably, the simplest experiment would measure the
spatial patterns of APD or Ca alternans on a fiber under various pacing
protocols. It is well know that in vitro experiments suffer nonstationary
drifts due to tissue dying. However, this is a slower process compared to
typical experimental protocols, which last a few tens of minutes. Moreover,
time constant of cardiac tissue is on the order of a few tens of seconds
\cite{Kalb2004JCardiovasc}. Therefore, it is possible to conduct multiple
well-designed protocols in a time interval when tissue dynamics remains
stationary. Other related experiments include the work of Fenton
\cite{Fenton2006KITP}, where action potential is measured using microelectrode
at a cell in a slice of paced dog epicardium tissue. Fenton showed that the
onset and the form of alternans at the measured cell depend on the pacing
history. While Fenton's observations may be attributable to factors such as
cardiac memory, the coexistence of multiple alternans patterns in cardiac
tissue may be another contributing element.

As pointed out by Qu and Weiss \cite{Qu2007AmJPHy}, because Ca$_{\text{in}}$
instability and V$_{\text{m}}$ instability are always being regulated by each
other, it remains a challenge to identify the sources of instabilities in
cardiac experiments. Recent work by Sato et al. \cite{Sato2007Biophys} and by
Jordan and Christini \cite{Jordan2007AmJPhy} have developed theoretical
criteria to assess the relative contributions of V$_{\text{m}}$ and
Ca$_{\text{in}}$ dynamics in inducing cardiac alternans. The coexistence of
multiple alternans patterns observed here may provide another possible
criterion for this purpose. If the phenomenon was verified in experiments, it
will raise questions in detection of alternans. New theories will also be
needed to understand the development of different alternans patterns to ensure
better prediction and control of spatiotemporal alternans.

\begin{center}
\textbf{Acknowledgments}
\end{center}

The author would like to thank David Schaeffer, Daniel Gauthier, Wanda
Krassowska, and Kevin Gonzales for their insightful comments. Particular
thanks go to Yohannes Shiferaw for providing a Fortran code of the
Shiferaw-Fox model and for answering many technical questions regarding the
model. The author is also grateful to Daisuke Sato and Alain Karma for their
useful discussions during the KITP miniprogram on cardiac dynamics. The author
would like to thank the anonymous reviewers for their thoughtful comments and
constructive suggestions. Support of the National Institutes of Health under
grant 1R01-HL-72831 is gratefully acknowledged.

\appendix

\section{Simulation of the Mapping Model}

Here, we present numerical examples for single cells and for fibers using the
mapping model (\ref{eqn:bcmap_A}, \ref{eqn:bcmap_C}). For simulation purpose,
we let $f\left(  D_{n}\right)  =\alpha\,\left(  1-\beta\,e^{-D_{n}/\tau
}\right)  $ and $g\left(  C_{n}\right)  =-s\,\left(  C_{n}-C_{\text{crit}%
}\right)  -C_{\min}$. Inspired by the results of Shiferaw et al. (Fig. 10 (d)
in \cite{Shiferaw2003BioPhy}), we choose the slope $s$ in $g$ to
be$\,\ s=\frac{1}{2}\left(  1-\theta\right)  \,s_{1}+\frac{1}{2}\left(
1+\theta\right)  \,s_{2}$, where $\theta=\tanh\left(  k\,\left(
C-C_{\text{crit}}\right)  \,\right)  $. To produce numerical results, we adopt
the following set of parameters: $\alpha=500$ ms, $\beta=0.62,$ $\tau=500$ ms,
$\gamma=-30$ ms/$\mu$M, $\mu=0.0025$ $\mu$M$/$ms, $C_{crit}=1.6$ $\mu$M,
$C_{\min}=1.25$ $\mu$M, $s_{1}=-2.5$, $s_{2}=0.01$, and $k=10\,\mu M^{-1}$.
\ Note that we have arbitrarily chosen a negative value of $\gamma$ although a
positive value of $\gamma$ will produce similar phenomena. Moreover, we make
the magnitude of $\gamma\,\mu$ small to simulate weak V$_{\text{m}}%
$/Ca$_{\text{in}}$ coupling.

\subsection{Alternans for single cells}

\begin{figure}[tbh]
\centering
\begin{tabular}
[c]{cc}%
\includegraphics[width=2.25in]{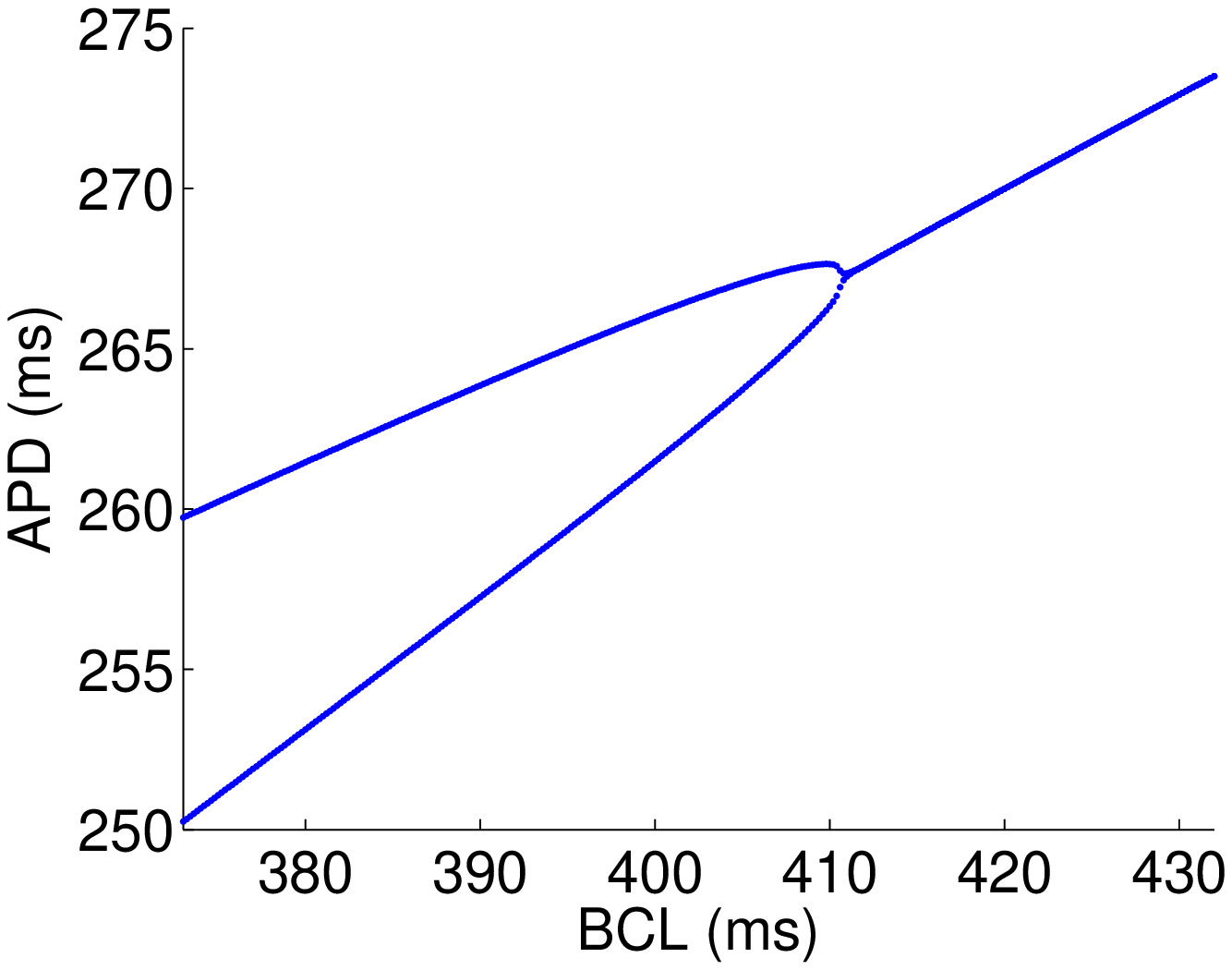} &
\includegraphics[width=2.25in]{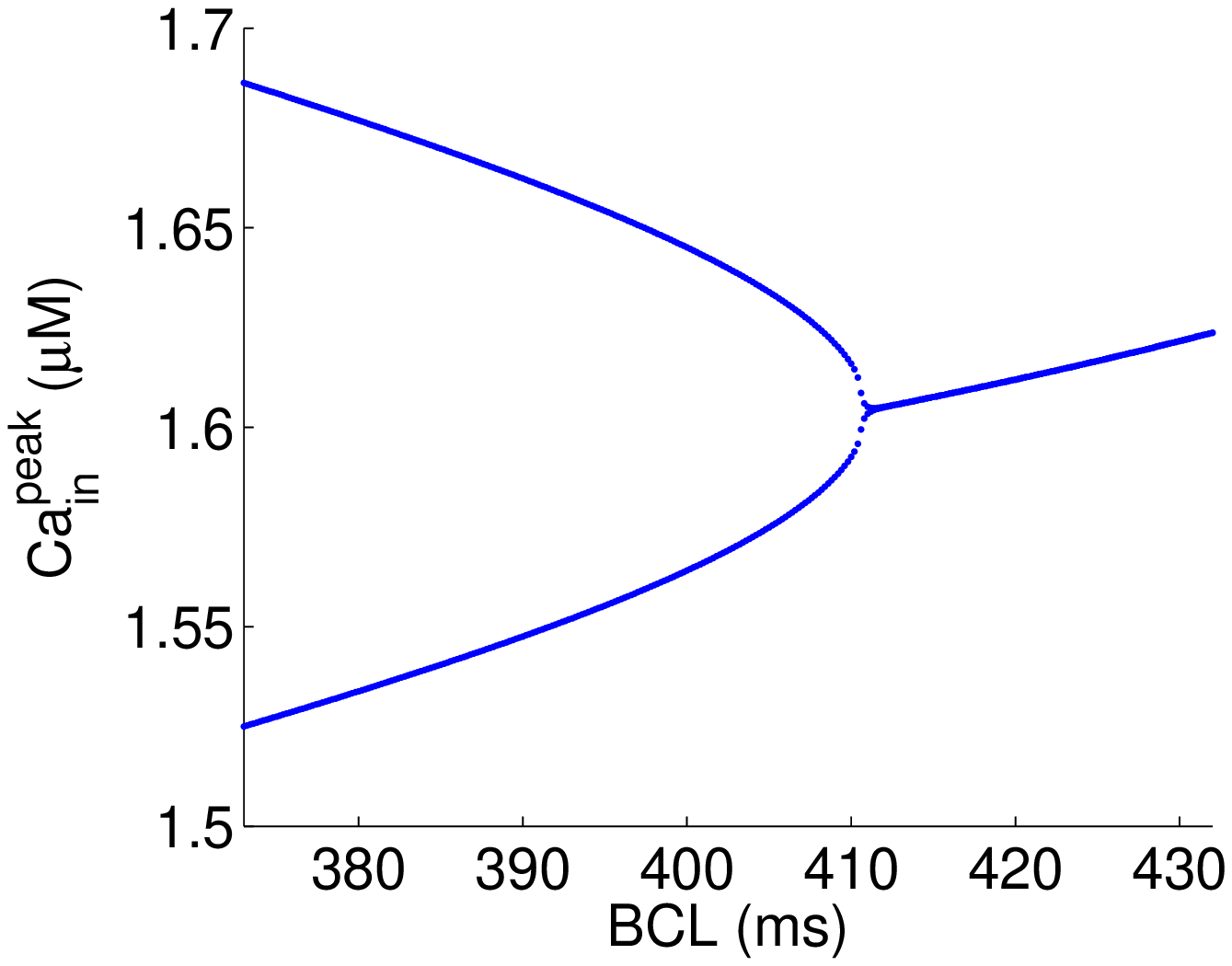}\\
(a) & (b)
\end{tabular}
\caption{Bifurcation diagrams of the map (\ref{eqn:bcmap_A}, \ref{eqn:bcmap_C}%
) with negative Vm/Ca coupling.}%
\label{fig:map_bifdiag}%
\end{figure}

With the chosen parameters, map (\ref{eqn:bcmap_A}, \ref{eqn:bcmap_C}) gives
birth to alternans through a period-doubling bifurcation at $B=410.958$ ms,
see the bifurcation diagrams in Fig. \ref{fig:map_bifdiag}. For this simple
model, it can be analytically shown that, for a given BCL, the cell has either
a unique 1:1 solution or a unique alternans solution. However, as we will see
below, a fiber has multiple alternans solutions induced from electrotonic
coupling. 

\subsection{Alternans for fibers}

Based on the mapping model (\ref{eqn:bcmap_A}, \ref{eqn:bcmap_C}) for single
cells, we construct a coupled-maps model to simulate fibers. Coupled-maps
models have been used by a few authors to study propagation of action
potential \cite{Fox2002CircRes,Fox2002PRL,Fox2003NewJPhy,Otani2007PRE}. Here,
we follow that approach. Specifically, we consider a fiber consisting of $M$
identical cells, located at $x_{i}$, $i=1,2,\ldots M$. The distance between
two neighboring cells is denoted by $\Delta x$. At cell $x_{i}$, APD and DI
are related by the following equation:
\begin{equation}
D_{n+1}\left(  x_{i}\right)  =T_{n+1}\left(  x_{i}\right)  -A_{n+1}\left(
x_{i}\right)  , \label{eqn:coupledmap_D}%
\end{equation}
where $T_{n+1}\left(  x_{i}\right)  $ is the time interval between two
consecutive activations of site $x_{i}$. The time $T_{n+1}\left(
x_{i}\right)  $ is determined by the propagation time from the pacing site to
$x_{i}$, that is,%
\begin{equation}
T_{n+1}\left(  x_{i}\right)  =B+\sum_{j=1}^{i-1}\frac{\Delta x}{V_{n+1}\left(
x_{i}\right)  }-\sum_{j=1}^{i-1}\frac{\Delta x}{V_{n}\left(  x_{i}\right)  },
\label{eqn:coupledmap_T}%
\end{equation}
where $V_{n}\left(  x_{i}\right)  $ stands for the conduction velocity. We
adopt a conduction velocity from Fox et al. \cite{Fox2003NewJPhy}:
$V_{n}\left(  x_{i}\right)  =V_{\max}\left(  1-\exp\left(  -\left(
D_{n}\left(  x_{i}\right)  +\beta\right)  /\delta\right)  \right)  $ with
parameters $V_{\max}=0.72$ cm/ms, $\beta=17.408,$ and $\delta=14$. To account
for the effect of electrotonic coupling, we modify the weighted averaging
formula in Fox et al. \cite{Fox2002CircRes} as follows:%
\begin{equation}
A_{n+1}\left(  x_{i}\right)  =\frac{\sum_{j=\max\left(  -30,1-i\right)
}^{\min\left(  30,\,M-i\right)  }w\left(  j\right)  \,\bar{A}_{n+1}\left(
x_{j}\right)  }{\sum_{j=\max\left(  -30,1-i\right)  }^{\min\left(
30,\,M-i\right)  }w\left(  j\right)  }, \label{eqn:coupledmap_A}%
\end{equation}
where $\bar{A}_{n+1}\left(  x_{j}\right)  =f\left(  D_{n}\left(  x_{j}\right)
\right)  +\gamma\,\left(  C_{n+1}\left(  x_{j}\right)  -C_{crit}\right)  $ and
$w\left(  j\right)  =\exp\left(  -0.0067\,j^{2}\right)  $. The equation for
[Ca$_{\text{in}}$] reads%
\begin{equation}
C_{n+1}\left(  x_{i}\right)  =\mu\,D_{n}\left(  x_{i}\right)  +g\left(
C_{n}\left(  x_{i}\right)  \right)  . \label{eqn:coupledmap_C}%
\end{equation}

\begin{figure}[ptb]
\centering%
\begin{tabular}{ccc}
\includegraphics[width=2.in]{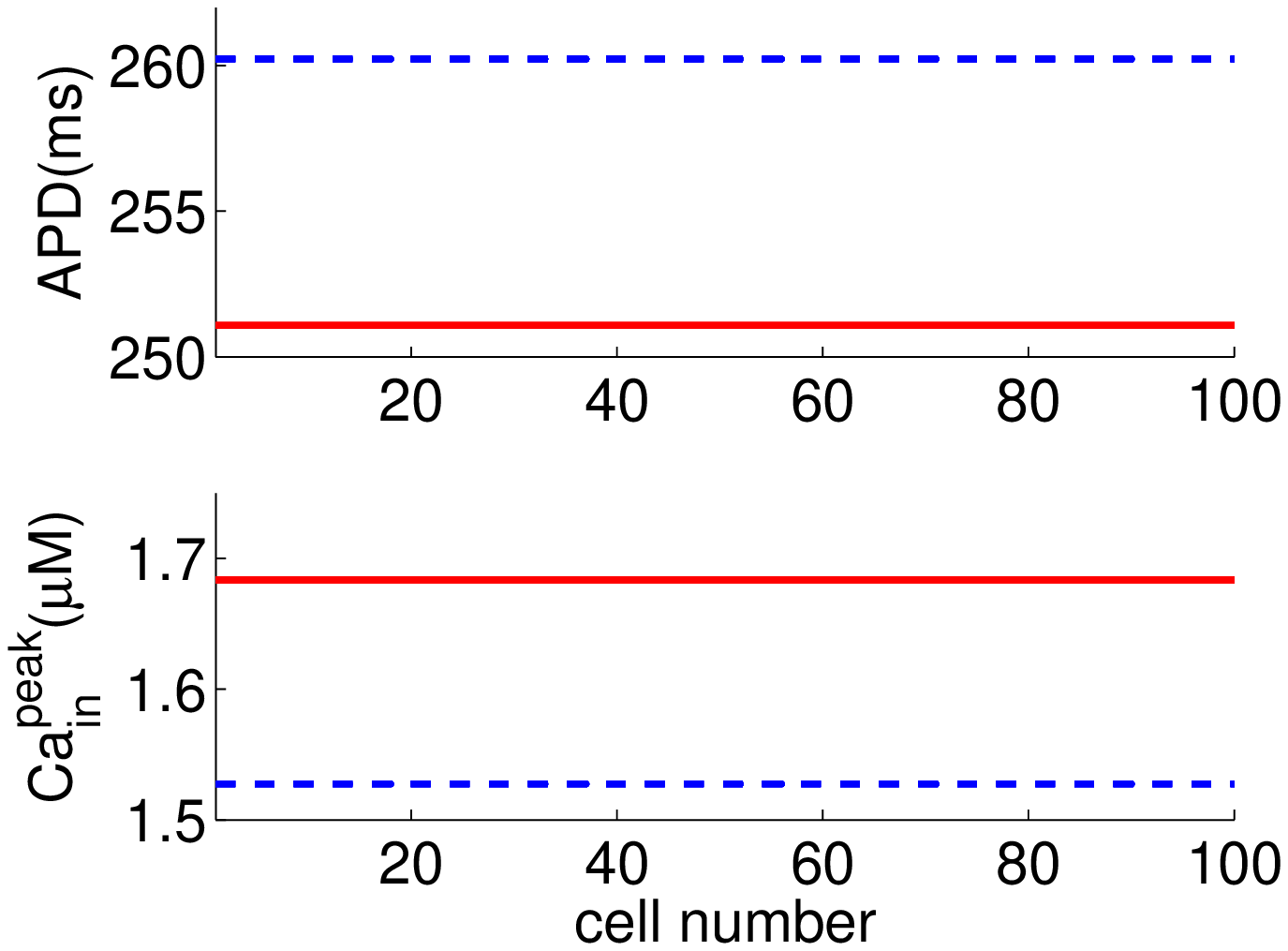} & %
\includegraphics[width=2.in]{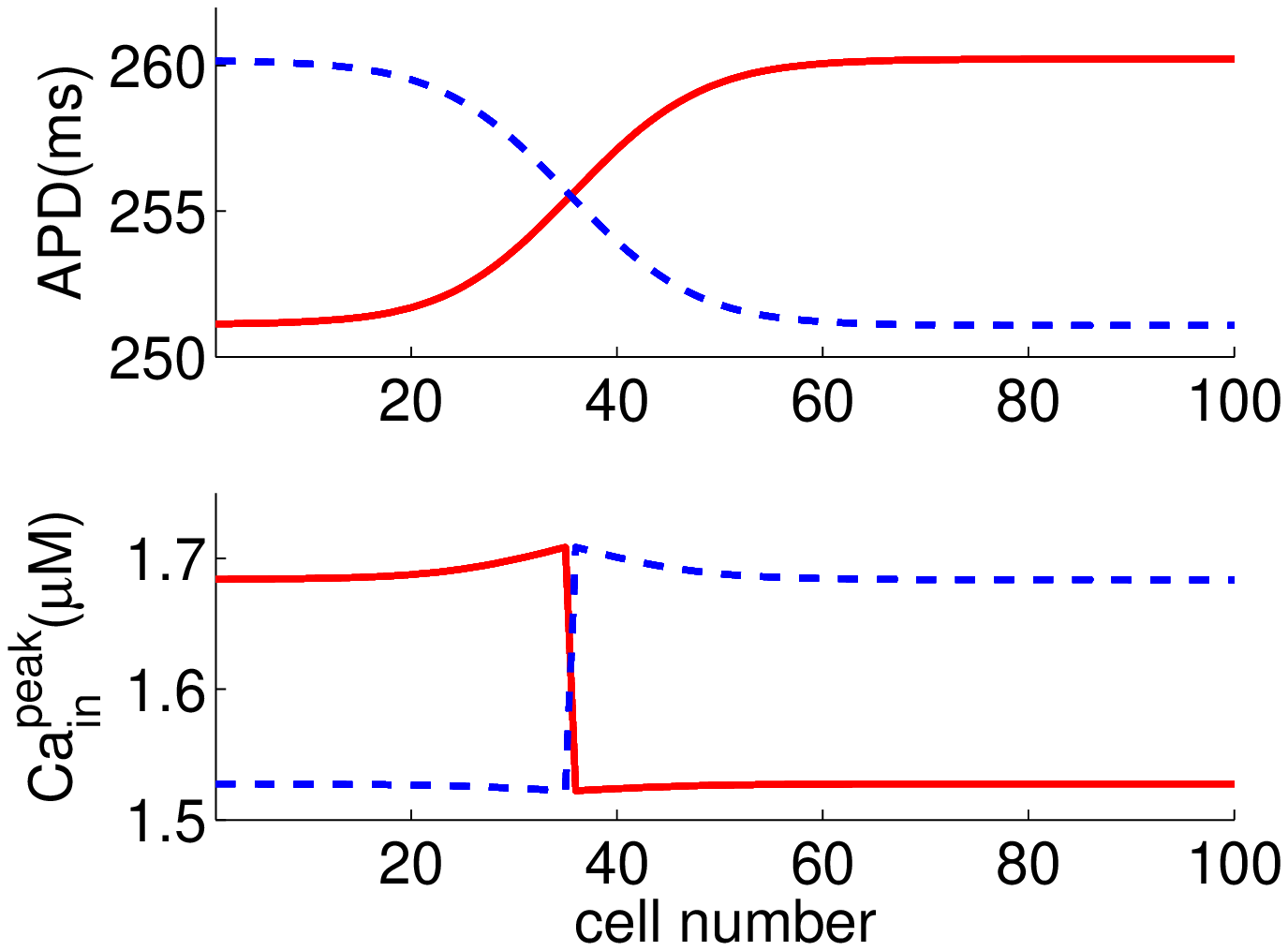} & %
\includegraphics[width=2.in]{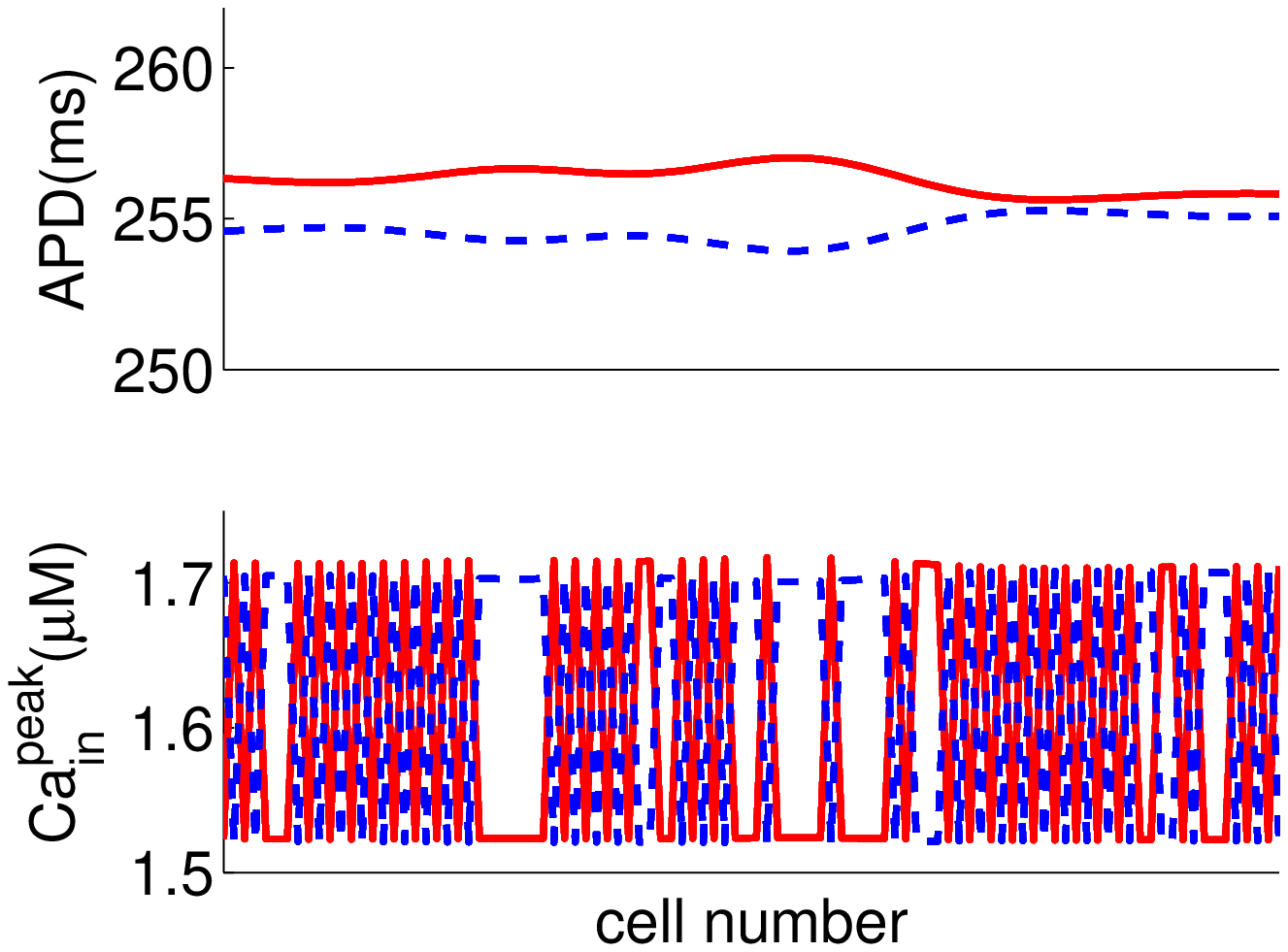} \\
(a) uniform distribution & (b) uneven distribution & (c) random distribution%
\end{tabular}%
\caption{Simulation of the coupled-maps model: three selected
alternans solutions for a homogeneous fiber of 100 cells with
negative Ca$\rightarrow V_{m}$ coupling. The pacing period is $375$
ms in all panels. Panel (a) starts from a uniform initial
distribution in Ca$_{\text{in}}$. Panel (b)
starts from an initial condition, where Ca$_{\text{in}}$ is set to be 1.5 $%
\protect\mu $M in the first 35 cells and 1.7 $\protect\mu $M in the
remaining cells. Panel (c) starts from a random initial distribution
in Ca. The last 10 beats of the steady-state solutions are plotted,
where odd beats are represented by red solid lines and even beats by
blue dashed lines. } \label{fig:negmap_100cells}
\end{figure}

Simulations of the coupled-maps model (\ref{eqn:coupledmap_D}%
-\ref{eqn:coupledmap_C}) show that the alternans pattern on a fiber depends on
the pacing history, a phenomenon consistent with that observed in simulations
of the Shiferaw-Fox model. For example, Fig. \ref{fig:negmap_100cells} shows 3
selected solutions for a homogeneous fiber of 100 cells paced at B=$375$ ms.
Panels in Fig. \ref{fig:negmap_100cells} have different initial conditions in
[Ca$_{\text{in}}$]. Panel (a) starts from a uniform initial distribution.
Panel (b) starts from an initial condition, where [Ca$_{\text{in}}$] is set to
be 1.5 $\mu$M in the first 35 cells and 1.7 $\mu$M in the remaining cells.
Panel (c) starts from a random initial distribution. Thus, the coupled maps
model is able to reproduce the coexistence of multiple alternans solutions as
observed in the Shiferaw-Fox model.

\begin{thebibliography}{99}                                                                                               %


\bibitem {AHA}\url{http://www.americanheart.org/}.

\bibitem {Karma1994}A. Karma, `Electrical alternans and spiral wave breakup in
cardiac tissue', Chaos, \textbf{4}, 461--472, 1994.

\bibitem {Garfinkel2000}A. Garfinkel, Y.H. Kim, O. Voroshilovsky, Z. Qu, J.R.
Kil, M.H. Lee, H.S. Karagueuzian, J.N. Weiss, and P.S. Chen, `Preventing
ventricular fibrillation by flattening cardiac restitution,' Proc. Natl Acad
Sci USA, \textbf{97}:6061--6066, 2000.

\bibitem {Plonsey2000}R. Plonsey and R.C. Barr, \emph{Bioelectricity: A
Quantitative Approach}, Kluwer, New York, 2000.

\bibitem {Nolasco1968JAP}J.B. Nolasco and R.W. Dahlen, `A Graphic Method for
the Study of Alternation in Cardiac Action Potentials,' Journal of Applied
Physiology, \textbf{25}:191--196, (1968).

\bibitem {Panfilov1998}A. Panfilov, `Spiral breakup as a model of ventricular
fibrillation,' Chaos \textbf{8}:57--64, 1998.

\bibitem {Pastore1999Circulation}J.M Pastore, S.D. Girouard, K.R. Laurita,
F.G. Akar, D.S. Rosenbaum, `Mechanism linking T-wave alternans to the genesis
of cardiac fibrillation,' Circulation. \textbf{99}:1385--1394, 1999.

\bibitem {Bloomfield2006ACC}D. M. Bloomfield, J. T. Bigger, R. C. Steinman, P.
B. Namerow, M. K. Parides, A. B. Curtis, E. S. Kaufman, J. M. Davidenko, T. S.
Shinn and J. M. Fontaine, `Microvolt T-wave alternans and the risk of death or
sustained ventricular arrhythmias in patients with left ventricular
dysfunction,' J. Am. Coll. Cardiol., \textbf{47}:456--463, 2006.

\bibitem {shusterman2006Cir}V. Shusterman, A. Goldberg and B. London, `Upsurge
in T-wave alternans and nonalternating repolarization instability precedes
spontaneous initiation of ventricular tachyarrhythmias in human,' Circulation,
\textbf{113}:2880--2887, 2006.

\bibitem {Shiferaw2006Review}Y. Shiferaw, Z. qu, A. Garfinkel, A. Karma, and
J.N. Weiss, `Nonlinear Dynamics of Paced Cardiac Cells,' Ann. N.Y. Acad. Sci.
\textbf{1080}:376--394, 2006.

\bibitem {Weiss2006CircRes}J.N. Weiss, A. Karma, Y. Shiferaw, P-S, Chen, A.
Garfinkel, and Z. Qu, `From Pulsus to Pulseless: The Saga of Cardiac
Alternans,' \textbf{98}:1244--1253, 2006.

\bibitem {EndoPhysiol1977}M. Endo, `Calcium release from the sarcoplasmic
reticulum,' Physiol Rev \textbf{57}:71--108, 1977.

\bibitem {Shiferaw2005PRE}Y. Shiferaw, D. Sato, and A. Karma, `Coupled
dynamics of voltage and calcium in paced cardiac cells,' Phys Rev E,
\textbf{71}:021903, 2005.

\bibitem {Shiferaw2006PNAS}Y. Shiferaw and A. Karma, `Turing Instability
mediated by voltage and calcium diffusion in paced cardiac cells,' PNAS,
\textbf{103}:5670--5675, 2006.

\bibitem {Sato2006CircRes}D. Sato, Y. Shiferaw, A. Garfinkel, J.N. Weiss, Z.
Qu, and A. Karma, `Spatially Discordant Alternans in Cardiac Tissue: Role of
Calcium Cycling,' Circulation Research, \textbf{99}:520--527, 2006.

\bibitem {Chudin1999Biophys}E. Chudin, J. Goldhaber, A. Garfinkel, J. Weiss,
and B. Kogan, `Intracellular Ca(2+) dynamics and the stability of ventricular
tachycardia,' Biophysics Journal, \textbf{77}:2930--2941, 1999.

\bibitem {Diaz2004CircRes}M.E. D\'{\i}az, S.C. O'Neill and D.A. Eisner,
`Sarcoplasmic Reticulum Calcium Content Fluctuation Is the Key to Cardiac,'
Circulation Research, \textbf{94}:650--656, 2004.

\bibitem {Rubenstein1995Circ}D.S. Rubenstein and S.L. Lipsius, Circulation
\textbf{91}, 201 (1995).

\bibitem {Walker2003CardioRes}M.L. Walker and D.S. Rosenbaum, Cardiovasc. Res.
\textbf{57}, 599 (2003).

\bibitem {Echebarria2002PRL}B. Echebarria and A. Karma, `Instability and
Spatiotemporal Dynamics of Alternans in Paced Cardiac Tissue,' Phys. Rev.
Lett. \textbf{88}:208101, 2002.

\bibitem {Echebarria2006}B. Echebarria and A. Karma, `Amplitude equation
approach to spatiotemporal dynamics of cardiac alternans,' to appear in PRE.

\bibitem {Dai2008}S. Dai and D. G. Schaeffer, `Spectrum of a linearized
amplitude equation for alternans in a cardiac fiber,' in process.

\bibitem {Choi2000JPhysiol}B.R. Choi, G. Salama, `Simultaneous maps of optical
action potentials and calcium transients in guinea-pig hearts: mechanisms
underlying concordant alternans,' J Physiol. \textbf{529}:171--188, 2000.

\bibitem {Pruvot2004CircRes}E.J. Pruvot, R.P. Katra, D.S Rosenbaum, K.R.
Laurita, `Role of calcium cycling versus restitution in the mechanism of
repolarization alternan,' Circ Res. \textbf{94}:1083--1090, 2004.

\bibitem {Katra2004HeartCircPhy}R.P. Katra, E. Pruvot, K.R. Laurita,
`Intracellular calcium handling heterogeneities in intact guinea pig hearts,'
Am J Physiol Heart Circ Physiol, \textbf{286}:H648--H656, 2004.

\bibitem {Aistrup2006CircRes}G.L. Aistrup, J.E. Kelly, S. Kapur, M. Kowalczyk,
I. Sysman-Wolpin, A.H. Kadish, and J.A. Wasserstrom, 'Pacing-induced
Heterogeneities in Intracellular Ca Signaling, Cardiac Alternans, and
Ventricular Arrhythmias in Intact Rat Heart,' Circulation Research,
\textbf{99}:65--73, 2006.

\bibitem {Shiferaw2003BioPhy}Y. Shiferaw, M. Watanabe, A. Garfinkel, J. Weiss,
A. Karma, `Model of intracellular calcium cycling in ventricular myocytes,'
Biophys J., \textbf{85}:3666--3686, 2003.

\bibitem {Fox2002AmJP}J. J. Fox, J. L. McHarg, and R. F. Gilmour, Am. J.
Physiol., `Ionic mechanism of electrical alternans,' \textbf{282}:H516--H530, 2002.

\bibitem {ShiferawCode}A Fortran code of the Shiferaw-Fox model was made by
Yohannes Shiferaw and is available at \url{http://www.csun.edu/~yshiferaw/shiferaw.html}.

\bibitem {Kalb2004JCardiovasc}S.S. Kalb, H.M. Dobrovolny, E.G. Tolkacheva,
S.F. Idriss, W. Krassowska and D.J. Gauthier, `The restitution portrait: a new
method for investigating rate-dependent restitution,' J. Cardiovasc.
Electrophysiol., \textbf{15}:698--7.

\bibitem {Riccio1999CircRes}M.L. Riccio, M.L., Koller, R.F. Gilmour Jr.,
`Electrical restitution and spatio-temporal organization during ventricular
fibrillation,' Circulation Research, \textbf{84}:955--963, 1999.

\bibitem {KrassowskaEqns}W. Krassowska, unpublished.

\bibitem {CaCoupling}In models without junctional diffusion, [Ca$_{\text{in}}%
$] in neighboring cells are indirectly coupled through their dependence on the voltage.

\bibitem {Bers2001}D.M. Bers, \emph{Excitation-Contraction Coupling and
Cardiac Contractile Force (Developments in Cardiovascular Medicine)}, Kluwer,
Boston, 2001.

\bibitem {Allbritton1992Science}N.L. Allbritton; T. Meyer, and L. Stryer,
`Range of Messenger Action of Calcium Ion and Inositol 1,4,5-Trisphosphate,'
Science, \textbf{258}:1812--1815, 1992.

\bibitem {SatoCommunication}Yohannes Shiferaw, Daisuke Sato, and Alain Karma,
private communication.

\bibitem {Nayfeh1973Perturbation}A.H. Nayfeh, \emph{Perturbation Methods},
John Wiley Interscience, New York, 1973.



\bibitem {Fox2002CircRes}J.J. Fox, M.L. Riccio, F. Hua, E. Bodenschatz, and
R.F. Gilmour Jr., `Spatiotemporal Transition to Conduction Block in Canine
Ventricle,' Circulation Research, \textbf{90}:289--296, 2002.

\bibitem {Courtemanche1993PRL}M. Courtemanche, L. Glass, and J.P. Keener,
`Instabilities of a Propagating Pulse in a Ring of Excitable Media,' Physical
Review Letters, \textbf{70}:2182--2185, 1993.

\bibitem {BeelerJPhysio1977}G.W. Beeler and H. Reuter, Reconstruction of the
Action Potential of Ventricular Myocardial Fibres, J. Physiol. \textbf{268}%
:177-210, 1977.

\bibitem {Cherry2007AmJPhy}E.M. Cherry and F.H. Fenton, `A Tale of Two Dogs:
Analyzing Two Models of Canine Ventricular Electrophysiology,' Am J Physiol
Heart Circ Physiol, \textbf{292}:H43-H55, 2007.

\bibitem {Hund2004Cir}T.J. Hund and Y. Rudy, `Rate Dependence and Regulation
of Action Potential and Calcium Transient in a Canine Cardiac Ventricular Cell
Model,' Circulation \textbf{110}: 3168-3174, 2004.

\bibitem {Fenton2006KITP}F. Fenton, `Beyond Slope One: Alternans Suppression
and Other Understudied Properties of APD Restitution,' KITP Miniprogram on
Cardiac Dynamics, Kavli Institute for Theoretical Physics, Santa Barbara, CA,
July 28, 2006.

\bibitem {Qu2007AmJPHy}Z. Qu and J.N. Weiss, `The chicken or the egg? Voltage
and calcium dynamics in the heart,' Am J Physiol Heart Circ Physiol,
\textbf{293}:H2054--H2055, 2007.

\bibitem {Sato2007Biophys}D. Sato, Y. Shiferaw, Z. Qu, A. Garfinkel, J.N.
Weiss, A. Karma, `Inferring the cellular origin of voltage and calcium
alternans from the spatial scales of phase reversal during discordant
alternans,' Biophys J, \textbf{92}:L33--L35, 2007.

\bibitem {Jordan2007AmJPhy}P.N. Jordan, D.J. Christini, `Characterizing the
contribution of voltage- and calcium-dependent coupling to action potential
stability: implications for repolarization alternans,' Am J Physiol Heart Circ
Physiol, \textbf{293}:H2109--H2118, 2007.

\bibitem {Fox2002PRL}J.J. Fox, R.F. Gilmour Jr., and E. Bodenschatz,
Conduction Block in One-Dimensional Heart Fibers, PRL, \textbf{89}:198101-1, 2002.

\bibitem {Fox2003NewJPhy}J.J. Fox, M.L. Riccio, P. Drury, A. Werthman, and
R.F. Gilmour Jr., `Dynamic mechanism for conduction block in heart tissue,'
New Journal of Physics, \textbf{5}:101.1--101.14, 2003.09, 2004.

\bibitem {Otani2007PRE}N.F. Otani, `Theory of action potential wave block
at-a-distance in the heart,' PRE, \textbf{75}, 021910, 2007.
\end{thebibliography}
\end{document}